\documentclass[journal,12pt,one column]{IEEEtran}

\usepackage{xspace,amsmath,amssymb,epsfig,syntonly,color, cite}
\usepackage{subfigure}
\newtheorem{lemma}{\hspace{-11pt}\bf Lemma}
\newtheorem{algorithm}{\hspace{-11pt}\bf Algorithm}

\newtheorem{proposition}{\hspace{-11pt}\bf Proposition}

\newtheorem{remark}{\hspace{-11pt}\bf Remark}

\long\def\symbolfootnote[#1]#2{\begingroup
\def\thefootnote{\fnsymbol{footnote}}
\footnote[#1]{#2}\endgroup}

\psfull

\begin{document}
\title{Beamforming Design for Multiuser Two-Way Relaying: A Unified Approach via Max-Min SINR}

\author{
   {Zhaoxi~Fang},
   {Xin~Wang (contact author)},
   {and Xiaojun~Yuan}\\
    \thanks{Z. Fang is with Dept. of Elec.\& Info. Engineering,
Zhejiang Wanli University, Ningbo, China, email:
zhaoxifang{\rm\char64}gmail.com; X. Wang is with Dept. of
Communication Science and Engineering, Fudan University, Shanghai,
China, email: xwang11{\rm\char64}fudan.edu.cn; X. Yuan is with the
Institute of Network Coding, Dept. of Information Engineering, The
Chinese University of Hong Kong, e-mail:
xjyuan{\rm\char64}inc.cuhk.edu.hk. } }

\maketitle 

\vspace*{-2.0cm}
\begin{abstract}
In this paper, we develop a unified framework for beamforming designs in
non-regenerative multiuser two-way relaying (TWR). The
core of our framework is the solution to the max-min
signal-to-interference-plus-noise-ratio (SINR) problem for multiuser TWR. We solve this problem
using a Dinkelbach-type algorithm with near-optimal performance and superlinear convergence.
We show that, using the max-min SINR solution as a corner stone, the beamforming designs under various important criteria, such as
weighted sum-rate maximization, weighted sum mean-square-error (MSE) minimization, and average bit-error-rate (BER) or symbol-error-rate (SER) minimization, etc, can be reformulated into a monotonic program. A polyblock outer approximation algorithm is then
used to find the desired solutions with guaranteed convergence and optimal performance (provided that the core max-min SINR solver is optimal). Furthermore, the proposed unified approach can provide important insights for tackling the optimal beamforming designs in other
emerging network models and settings. For instances, we extend the proposed framework to address the beamforming design in collaborative TWR
and multi-pair MIMO TWR. Extensive numerical results are presented to demonstrate the merits of the proposed beamforming
solutions.

\vspace*{-0.1cm}
\end{abstract}
\vspace*{-0.8cm}
   {\small
      \begin{center}
        \[
           \begin{array}{rl}
               \text{\bf Submission date:} & \text{March 24, 2013}\\
               \text{\bf Revision Date:} & \text{\today} \\
               \text{\bf Associate Editor:} & \text{Prof. Eduard Jorswieck} \\
           \end{array}
         \]
      \end{center}
    }

\markboth{}{}


\section{Introduction}

Relay communications have been long studied to enhance the capacity
and expand the coverage of wireless networks. For conventional
communications between two users via a single relay, four transmission
phases in time or frequency are typically required: two used for
user-to-relay, and the other two for relay-to-user. To improve
spectral efficiency, a two-way relaying (TWR) method, referred to as
physical-layer network coding (PNC) \cite{Zha06}, was proposed
to accomplish bidirectional data exchange in two phases. This PNC technique is remarkable for its potential to double the system throughput.

PNC for two-way relay channels has gained a growing interest in recent
years \cite{Zha06,Ale09,Kat07,NazerIT11}. Various relaying
strategies have been proposed to exploit the benefit of PNC,
including but not limited to, decode-and-forward \cite{Zha06},
compress-and-forward \cite{Ale09}, amplify-and-forward (AF)
\cite{Kat07}, and compute-and-forward \cite{NazerIT11}.
Particularly, it was shown in \cite{NamIT10} that PNC with nested
lattice coding can achieve the capacity of the single-input
single-output Gaussian two-way relay channel within $\frac{1}{2}$
bit. Later, the authors in \cite{Yang12,Yang13} showed that
lattice-coding techniques can be efficiently incorporated into
multiple-input multiple-output (MIMO) TWR, where the users and the
relay are equipped with multiple antennas. It was revealed therein that
near-capacity performance can be achieved in MIMO two-way relay channels.


More recently, multiuser two-way relaying, in which multiple
users exchange data via a single relay in a pairwise or non-pairwise
manner, has been intensively studied in the literature \cite{Che09,Jou10,WangF12,ZBRHP12,Zha09,Tao12,ZRH12,WangF13}.
In these approaches, analogue network coding (ANC) is employed, i.e.,
simple AF operations are implemented at the relay and self interference
is canceled at the user ends;
multiple antennas are deployed at the relay to provide
extra degrees of freedom, which enables a potential boost of the system
throughput. However, to fully exploit this potential requires a proper
design of the beamforming (or called \emph{precoding}) matrix at the
relay, which is in general a difficult problem. To date, only approximate algorithms
have been proposed based on specific design criteria, such as
zero-forcing \cite{WangF12}, power minimization \cite{ZBRHP12},
max-min signal-to-interference-plus-noise ratio
(SINR) \cite{Tao12}, and maximum sum-rate \cite{ZRH12,WangF13}.



In this paper, we develop a unified framework to solve
the beamforming optimization problems for multiuser TWR. We use the classic max-min SINR
problem as the core of our framework. Our major contribution
is to show that the max-min SINR solution can be used as a
corner stone to pursue the optimal beamforming designs based
on arbitrary utility functions that are monotonic in the user
SINRs. Our framework works for various optimization criteria,
such as power minimization, weighted sum-rate maximization,
average symbol-error-rate (SER) or bit-error rate (BER)
minimization, etc. Relying on solving a series of max-min
SINR problems, a polyblock outer approximation algorithm
is developed to find the desired solutions with guaranteed
convergence and global optimality (provided that the core
max-min SINR solver yields the optimal solution).

The optimality and efficiency of our proposed framework depends on the
choice of the max-min SINR solver. In our approach, the max-min SINR problem, treated as a max-min fractional
program, is solved using a Dinkelbach-type algorithm \cite{Fre06}. This algorithm is optimal for the two-user case and can
provide near-optimal performance for the general case of multiple pairs of users. It is worth mentioning that the max-min SINR problem can be alternatively solved using the bisection search method in \cite{Tao12} with linear (i.e., geometrically fast) convergence. In contrast, the proposed Dinkelbach-type algorithm has a quotient- (Q-)superlinear convergence speed \cite{Fre06}, and hence in general exhibit
faster convergence (and thus reduced computation) than the bisection search method.

Furthermore, the proposed unified approach can provide important insights for tackling the optimal beamforming designs in other
emerging network models and settings. For instances, we extend the proposed framework to cover the beamforming design in collaborative TWR
and multi-pair MIMO TWR. Specifically, for collaborative TWR, we propose the beamforming
design under an individual power constraint at each relay node, which is more practical than
the settings in \cite{ZHKV12,CGHMW11} (where the relays share a total power budget). For multi-pair MIMO
TWR, an iterative optimization algorithm is developed to jointly optimize the transmit and
receive beamforming vectors of each user, together with the relay precoding matrix.
Extensive numerical results are presented to demonstrate the merits of the proposed beamforming
solutions.

The rest of this paper is organized as follows. Section II outlines
the notations in use and the system model. Section III discusses the
max-min SINR problem and its solution, as well as the relation
between the power minimization design and the max-min SINR design. A
unified approach for beamforming designs is presented in Section IV.
Sections V and VI discuss generalizations of the proposed framework
to collaborative beamforming for multi-pair multi-relay TWR, as well
as to multi-pair MIMO TWR. The proposed schemes are tested and
compared with existing alternatives in Section VII, followed by the
conclusions in Section VIII.


\section{Preliminaries}

\subsection{Notation}

The following notation is used throughout this paper. Boldface fonts
denote vectors or matrices, the $i$th entry of a vector, say
$\boldsymbol{a}$, is denoted by $a_i$; 
$\mathbb{R}^{K\times M}$ and $\mathbb{C}^{K\times M}$ denote the
$K$-by-$M$ dimensional real and complex space, respectively.
$\mathbb{R}_+^K : = \{\boldsymbol{a} \in \mathbb{R}^{K \times 1}
\;|\; \boldsymbol{a} \geq \boldsymbol{0}\}$. Note that the vector
inequalities, such as $\boldsymbol{a} \geq \boldsymbol{0}$, are defined element-wise. $ \left \lceil x \right
\rceil$ denotes the nearest integer greater than or equal to $x$;
$(\cdot)^*$ denotes complex conjugate, $(\cdot)^T$ denotes
transpose, and $(\cdot)^H$ conjugate transpose; $\otimes$ represents
the Kronecker product; $\odot$ denotes the Schur-Hadamard
(element-wise) product; $\text{tr}(\boldsymbol{A})$ denotes trace
operator for matrix $\boldsymbol{A}$, $\text{vec}(\boldsymbol{A})$
operator creates a column vector from a matrix $\boldsymbol{A}$ by
stacking its column vectors below one another,
$\boldsymbol{A}^{1/2}$ denotes the square-root of a positive
semi-definite matrix $\boldsymbol{A}$,
$\text{diag}(\boldsymbol{A}_1, \ldots, \boldsymbol{A}_M)$ denotes a
block-diagonal matrix with $\boldsymbol{A}_1, \ldots,
\boldsymbol{A}_M$ as the submatrices in the diagonal; $\| \cdot \|$
denotes the Euclidean norm for vectors, and $| \cdot |$ denotes norm
of a complex scalar; $\boldsymbol{0}$ and $\boldsymbol{1}$ denote
all-zero and all-one vectors; $\boldsymbol{A} \succeq 0$ means that
a square matrix $\boldsymbol{A}$ is positive semi-definite; a
circularly symmetric complex Gaussian random vector $\boldsymbol{x}$
with mean $\boldsymbol{\bar{x}}$ and covariance matrix
$\boldsymbol{\Sigma}$ is denoted as $\boldsymbol{x} \thicksim {\cal
CN}(\boldsymbol{\bar{x}},\boldsymbol{\Sigma})$, where $\thicksim$
stands for ``distributed as''; ${\cal A}\backslash {\cal B}$ denotes
the set obtained by excluding all the elements of set ${\cal B}$
from set ${\cal A}$.

\subsection{System Model for Multi-Pair TWR}

\begin{figure}
\centering \epsfig{file= 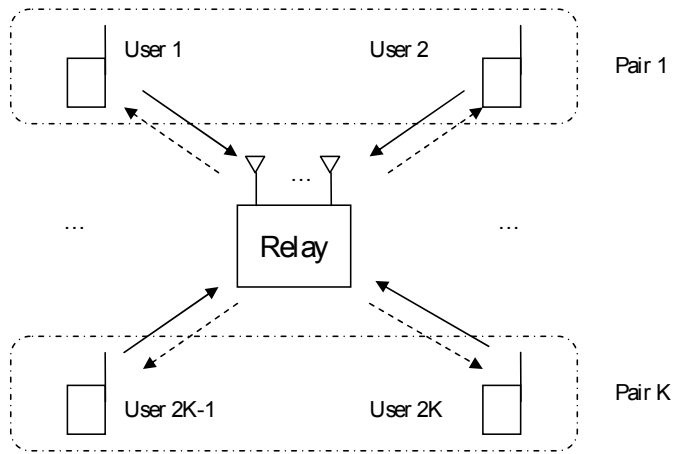, width=0.5\textwidth} \caption{ A
multi-pair two-way relaying system.} \label{F:fig1} \vspace{-0.3cm}
\end{figure}

As shown in Fig. 1, we consider a two-way relay (bidirectional)
communication between $K$ pairs of users, where the relay is
equipped with $M$ antennas and each user has a single antenna
\cite{Che09, Jou10}. Without loss of generality, it is assumed that
the $(2k-1)$th and the $(2k)$th users communicate with each other,
$k=1, \ldots, K$, through two phases. The communication channels
between the relay and users are assumed to be flat-fading over a
common narrow band. Following the convention in \cite{Che09,
Jou10,ZRH12,WangF13}, we assume global channel state information
(CSI), i.e., all the users and the relay have full CSI.

In the first phase of the two-way relaying
communication, all users transmit to the relay simultaneously, and
the received signal $\boldsymbol{y}_R(t) \in \mathbb{C}^{M \times
1}$ at the relay is
\begin{equation}\label{eq.yR}
    \boldsymbol{y}_R(t) = \sum_{i=1}^{2K} \boldsymbol{h}_i \sqrt{p_i} s_i(t) + \boldsymbol{n}_R(t),
\end{equation}
where $\boldsymbol{h}_i$, $p_i$ and $s_i(t)$ denote the channel coefficient vector from user $i$ to the relay, transmit power of user $i$, and unit-power transmitted symbol from user $i$, respectively, and $\boldsymbol{n}_R(t) \in \mathbb{C}^{M \times 1}$ denotes the noise vector. With a given covariance matrix $\boldsymbol{\Lambda}_R$, it is assumed $\boldsymbol{n}_R(t) \thicksim {\cal CN}(\boldsymbol{0},\boldsymbol{\Lambda}_R)$.

Upon receiving $\boldsymbol{y}_R(t)$, the non-regenerative relay amplifies
and forwards the signal $\boldsymbol{x}_R(t) = \boldsymbol{A}
\boldsymbol{y}_R(t)$ to all users in the next phase, where $\boldsymbol{A} \in \mathbb{C}^{M \times M}$ is the relay beamforming
matrix. The transmit power at the relay is
\begin{equation}
\begin{split}
p_R(\boldsymbol{A}) &= E\|\boldsymbol{x}_R(t)\|^2 \nonumber \\
&=  E\left\| \boldsymbol{A} (\sum_{i=1}^{2K} \boldsymbol{h}_i \sqrt{p_i} s_i(t) + \boldsymbol{n}_R(t))\right\|^2 \nonumber \\
&= \sum_{i=1}^{2K} p_i \|\boldsymbol{A}\boldsymbol{h}_i\|^2 + \text{tr}(\boldsymbol{A} \boldsymbol{\Lambda}_R \boldsymbol{A}^H).
\end{split}
\end{equation}

Suppose that channel reciprocity holds for the uplink and downlink
transmission between the relay and users. The received signal at
user $i\in\{1,\ldots, 2K\}$ is given by
\begin{equation}\label{eq.yi}
    y_i(t) = \boldsymbol{h}_i^T \boldsymbol{A} \sum_{j=1}^{2K} \boldsymbol{h}_j \sqrt{p_j} s_j(t) + \boldsymbol{h}_i^T \boldsymbol{A} \boldsymbol{n}_R(t) + n_i(t)
\end{equation}
where the receive noise $n_i(t) \thicksim {\cal CN}(0, \sigma_i^2)$.

Upon receiving the downlink signal, user $(2k-1)$ intends to detect the signal $s_{2k}(t)$ from
user $2k$, and the term $\sqrt{p_{2k-1}} \boldsymbol{h}_{2k-1}^T
\boldsymbol{A} $ $ \boldsymbol{h}_{2k-1} s_{2k-1}(t)$ in
(\ref{eq.yi}) is referred to as ``self-interference''. In the spirit
of ANC, this self-interference can be canceled before signal
detection. The SINR at the $(2k-1)$th user is thus
\begin{equation}\label{eq.sinr1}
   \text{SINR}_{2k-1}(\boldsymbol{A})= \frac{p_{2k} |\boldsymbol{h}_{2k-1}^T \boldsymbol{A} \boldsymbol{h}_{2k}|^2}{\sum_{i \neq 2k-1, 2k} [p_i |\boldsymbol{h}_{2k-1}^T \boldsymbol{A} \boldsymbol{h}_i|^2]+\|\boldsymbol{\Lambda}_R^{1/2} \boldsymbol{A}^H \boldsymbol{h}_{2k-1}^*\|^2+ \sigma_{2k-1}^2};
\end{equation}
and, similarly, the SINR at the $(2k)$th user is
\begin{equation}\label{eq.sinr2}
    \text{SINR}_{2k}(\boldsymbol{A})=  \frac{p_{2k-1} |\boldsymbol{h}_{2k}^T \boldsymbol{A} \boldsymbol{h}_{2k-1}|^2}{\sum_{i \neq 2k-1, 2k} [p_i |\boldsymbol{h}_{2k}^T \boldsymbol{A} \boldsymbol{h}_i|^2]+\|\boldsymbol{\Lambda}_R^{1/2} \boldsymbol{A}^H \boldsymbol{h}_{2k}^*\|^2+ \sigma_{2k}^2}.
\end{equation}

Based on the SINRs (\ref{eq.sinr1}) and (\ref{eq.sinr2}), we will develop a unified approach for beamforming designs in AF-based TWR under different criteria.

\section{SINR Balancing Optimization}

In this section, we describe two alternative forms of the SINR balancing problem. The effective solution to this problem will serve as a corner stone of our proposed framework.

\subsection{Max-Min SINR Problem}

We start with the first form of SINR balancing, i.e., the max-min SINR problem formulated as
\begin{equation}\label{eq.maxmin2k}
\begin{split}
\lambda^{\text{opt}} = & \max_{\boldsymbol{A}} \min_{i=1, \ldots, 2K} \frac{\text{SINR}_i(\boldsymbol{A})}{\gamma_i}\\
\text{s. t.}  ~~~& \sum_{i=1}^{2K} {p_i} \|\boldsymbol{A}\boldsymbol{h}_i\|^2 + \text{tr}(\boldsymbol{A} \boldsymbol{\Lambda}_R \boldsymbol{A}^H) \leq \check{P}_R
\end{split}
\end{equation}
where $\gamma_i$ denotes the SINR target for user $i$, and $\check{P}_R$ denotes the total power budget at the relay.

Relying on a semi-definite programming (SDP) based Dinkelbach-type
algorithm, this max-min SINR problem has been solved for one-pair
(i.e., $K=1$) TWR \cite{Wan12}. The problem has also been approximately solved using bisection search over the SDP relaxation solvers for related power minimization problems for the general case of $K$-pair users \cite{Tao12}. Here, we generalize
the Dinkelbach-type algorithm in \cite{Wan12} to approximately solve (\ref{eq.maxmin2k}) for the case of $K$-pair users. We show that the proposed algorithm is more efficient than the bisection search method.

We start with the following definitions:
\begin{equation}\label{eq.sdpdef}
    \boldsymbol{q}_{ji} := \text{vec}(\boldsymbol{h}_j \boldsymbol{h}_i^T)\ \mathrm{and} \  \boldsymbol{B}_i := \text{diag}(\boldsymbol{h}_i^T, \ldots,\boldsymbol{h}_i^T) \in \mathbb{C}^{M\times(MM)}
\end{equation}
where $\boldsymbol{h}_i^T$ is repeated by
$M$ times in $\boldsymbol{B}_i$.


Let $\boldsymbol{\Theta} := \sum_{i=1}^{2K} [p_i \boldsymbol{h}_i
\boldsymbol{h}_i^H] + \boldsymbol{\Lambda}_R$, and
$\boldsymbol{\Phi}:= (\boldsymbol{\Theta}^{1/2})^T \otimes
\boldsymbol{I}_M$. Further let $\boldsymbol{a} :=
\text{vec}(\boldsymbol{A})$, $\boldsymbol{X} :=
\boldsymbol{a}\boldsymbol{a}^H$, $\boldsymbol{E}_0 :=
\boldsymbol{\Phi}^H\boldsymbol{\Phi}$. Then we have the relay
transmit power:
\[
\sum_{i=1}^{2K} {p_i}
\|\boldsymbol{A}\boldsymbol{h}_i\|^2 + \text{tr}(\boldsymbol{A}
\boldsymbol{\Lambda}_R \boldsymbol{A}^H) = \text{tr}(\boldsymbol{A}
(\sum_{i=1}^{2K} {p_i}\boldsymbol{h}_i\boldsymbol{h}_i^H +
\boldsymbol{\Lambda}_R) \boldsymbol{A}^H) = \|\boldsymbol{\Phi}
\boldsymbol{a}\|^2 = \text{tr}(\boldsymbol{E}_0 \boldsymbol{X}).
\]
With (\ref{eq.sdpdef}), we also have $|\boldsymbol{h}_j^T
\boldsymbol{A} \boldsymbol{h}_i|^2 =
|\boldsymbol{q}_{ji}^T\boldsymbol{a}|^2$ and $\boldsymbol{h}_i^T
\boldsymbol{A}\boldsymbol{\Lambda}_R \boldsymbol{A}^H
\boldsymbol{h}_i^* = \|\boldsymbol{\Lambda}_R^{1/2} \boldsymbol{A}^H
\boldsymbol{h}_i^*\|^2 = \|\boldsymbol{\Lambda}_R^{1/2}
\boldsymbol{B}_i \boldsymbol{a}\|^2$. Hence, we have
\begin{equation}
    \frac{\text{SINR}_{2k-1}(\boldsymbol{a})}{\gamma_{2k-1}}=
    \frac{p_{2k} |\boldsymbol{q}_{2k-1,2k}^T\boldsymbol{a}|^2} { \gamma_{2k-1}( {\sum_{i \neq 2k-1, 2k} p_i | \boldsymbol{q}_{2k-1,i}^T \boldsymbol{a}|^2}  + \|\boldsymbol{\Lambda}_R^{1/2} \boldsymbol{B}_{2k-1} \boldsymbol{a}\|^2+
    \sigma_{2k-1}^2)}, \nonumber
\end{equation}
and
\begin{equation}
     \frac{\text{SINR}_{2k}(\boldsymbol{a})}{\gamma_{2k}}=
     \frac{p_{2k-1} |\boldsymbol{q}_{2k,2k-1}^T\boldsymbol{a}|^2}{\gamma_{2k}({\sum_{i \neq 2k-1, 2k} p_i | \boldsymbol{q}_{2k,i}^T \boldsymbol{a}|^2} + \|\boldsymbol{\Lambda}_R^{1/2} \boldsymbol{B}_{2k} \boldsymbol{a}\|^2+
     \sigma_{2k}^2)}. \nonumber
\end{equation}

Define  $\boldsymbol{E}_{2k-1}^{(1)} :=p_{2k}
\boldsymbol{q}_{2k-1,2k}^*\boldsymbol{q}_{2k-1,2k}^T$,
$\boldsymbol{E}_{2k-1}^{(2)} := \sum_{i \neq 2k-1,2k} [p_i
\boldsymbol{q}_{2k-1,i}^*\boldsymbol{q}_{2k-1,i}^T]+\boldsymbol{B}_{2k-1}^H
\boldsymbol{\Lambda}_R \boldsymbol{B}_{2k-1}$,
$\boldsymbol{E}_{2k}^{(1)} :=  p_{2k-1}
\boldsymbol{q}_{2k,2k-1}^*\boldsymbol{q}_{2k,2k-1}^T$, and
$\boldsymbol{E}_{2k}^{(2)} :=  \sum_{i \neq 2k-1,2k} [p_i
\boldsymbol{q}_{2k,i}^*\boldsymbol{q}_{2k,i}^T]+\boldsymbol{B}_{2k}^H
\boldsymbol{\Lambda}_R \boldsymbol{B}_{2k}$, for $k=1,\ldots, K$.

In terms of $\boldsymbol{X}$, let
\begin{equation}\label{eq.fgi}
f_i(\boldsymbol{X}):= \text{tr}(\boldsymbol{E}_i^{(1)} \boldsymbol{X})\ \mathrm{and} \
g_i(\boldsymbol{X}):= \text{tr}(\boldsymbol{E}_i^{(2)}\boldsymbol{X}) + \sigma_i^2,\ \mathrm{for}\ i=1, \ldots, 2K.
\end{equation}

Using $\boldsymbol{X}$ as the optimization
variable and dropping the constraint of
$\text{rank}(\boldsymbol{X})=1$, we can relax (\ref{eq.maxmin2k}) to
\begin{equation}\label{eq.relax}
\begin{split}
\tilde{\lambda}^{\text{opt}} = & \max_{\boldsymbol{X}} \min_{i=1, \ldots, 2K} \frac{f_i(\boldsymbol{X})}{\gamma_i g_i(\boldsymbol{X})}\\
\text{s. t.}  ~~~& \boldsymbol{X}\succeq 0, \quad \text{tr}(\boldsymbol{E}_0 \boldsymbol{X}) \leq \check{P}_R.
\end{split}
\end{equation}

The problem (\ref{eq.relax}) is a max-min fractional program, and
can be solved using a primal Dinkelbach-type algorithm \cite{Fre06}.
This algorithm is based on solving a sequence of the following
parametric optimization problems for $\lambda \leq
\lambda^{\text{opt}}$:
\begin{equation}\label{eq.maxminX2k}
\begin{split}
    & \max_{\boldsymbol{X}} \min_{i=1,\ldots,2K} f_i(\boldsymbol{X}) - \lambda \gamma_i g_i(\boldsymbol{X})\\
    & \text{s. t.}  ~~~ \boldsymbol{X}\succeq 0, \quad \text{tr}(\boldsymbol{E}_0 \boldsymbol{X}) \leq \check{P}_R.
\end{split}
\end{equation}

Let $\boldsymbol{E}_i := \boldsymbol{E}_i^{(1)} - \lambda \gamma_i
\boldsymbol{E}_i^{(2)}$, $i=1,\ldots, 2K$. The problem
(\ref{eq.maxminX2k}) becomes a convex SDP as
\begin{equation}\label{eq.sdp}
\begin{split}
& \min_{\boldsymbol{X}, \tau} ~ -\tau\\
\text{s. t.}  ~~~& \boldsymbol{X}\succeq 0, \quad \text{tr}(\boldsymbol{E}_0 \boldsymbol{X}) \leq \check{P}_R, 
\quad \text{tr}(\boldsymbol{E}_i \boldsymbol{X})- \lambda \gamma_i \sigma_i^2 \geq \tau, \quad i=1,\ldots, 2K.
\end{split}
\end{equation}
This SDP can be solved by the interior point method in polynomial
time \cite{Ger10}.

Relying on this SDP solution, we propose the following algorithm to
solve (\ref{eq.relax}):

\begin{quote}
\vspace{0.05 in}\hrule height0.1pt depth0.3pt \vspace{0.05 in}
\begin{algorithm}
{\it for max-min SINR problem}

{\bf Initialize}:  $\boldsymbol{A}^{0} = (\frac{\check{P}_R}{\sum_{i=1}^{2K} (p_i \|\boldsymbol{h}_i\|^2) +\text{tr}(\boldsymbol{\Lambda}_R)})^{1/2} \boldsymbol{I}$,
$\boldsymbol{X}^{(0)} =
\text{vec}(\boldsymbol{A}^{0})\text{vec}(\boldsymbol{A}^{0})^H$, and
$j=0$.

{\bf Repeat}: $j=j+1$,\\
    given $\boldsymbol{X}^{(j-1)}$, find $\lambda^{(j)} = \min_{i=1,\ldots, 2K} \frac{f_i(\boldsymbol{X}^{(j-1)})}{\gamma _i g_i(\boldsymbol{X}^{(j-1)})}$; \\
    given $\lambda^{(j)}$, solve (\ref{eq.sdp}) with SDP to obtain:
    $\boldsymbol{X}^{(j)} = \arg \max_{\boldsymbol{X}} \min_{i=1,\ldots,2K} [f_i(\boldsymbol{X}) - \lambda^{(j)} \gamma _i g_i(\boldsymbol{X})]$;

{\bf until} $\min_{i=1,\ldots, 2K} [f_i(\boldsymbol{X}^{(j)}) -
\lambda^{(j)} \gamma _i g_i(\boldsymbol{X}^{(j)})]\leq 0$.

{\bf Output}: $\tilde{\lambda}^{\text{opt}} = \lambda^{(j)}$, and $\boldsymbol{X}^{(j)}$ as the solution.
\end{algorithm}
\vspace{0.05 in} \hrule height0.1pt depth0.3pt \vspace{0.1 in}
\end{quote}

Algorithm 1 is a classic Dinkelbach-type algorithm \cite{Fre06}. In Problem (\ref{eq.relax}), it is clear that $0 <
g_i(\boldsymbol{X}) < \infty, \forall \boldsymbol{X}$, and
$\tilde{\lambda}^{\text{opt}}$ is finite. Hence, Condition 8.5 in
\cite{Fre06} holds. According to \cite[Theorem 8.7]{Fre06}, we
immediately have the following result.
\begin{lemma}
Algorithm 1 converges Q-superlinearly\footnote{Let $\tilde{\lambda}^{\text{opt}}$ denote the optimal value of problem (\ref{eq.relax}), and $\lambda^{(j)}$
the output value of the $j$-th iteration of Algorithm 1. We say that the sequence $\lambda^{(j)}$ converges Q-superlinearly to $\tilde{\lambda}^{\text{opt}}$ if $\lim_{j \to \infty} \frac{|\lambda^{(j+1)} - \tilde{\lambda}^{\text{opt}}|} {|\lambda^{(j)} - \tilde{\lambda}^{\text{opt}}|} = 0$.}
to the global optimal solution $\boldsymbol{X}^{\mathrm{opt}}$ for (\ref{eq.relax}).
\end{lemma}

\begin{remark}
We note that the max-min SINR problem can be alternatively solved
with the bisection search method in \cite{Tao12}. It is known that
the bisectional search has a linear, i.e., geometrically fast
convergence speed. In contrast, the proposed Dinkelbach-type
algorithm has quotient-superlinear convergence. Therefore,
the proposed algorithm in general exhibits a faster convergence
speed than the bisection search method in \cite{Tao12}. We further
remark that, the proposed Algorithm 1 is guaranteed to converge to the optimal solution of (\ref{eq.relax}) from any feasible initial $\boldsymbol{A}^{0}$ per Lemma 1. Here, we set $\boldsymbol{A}^{0}$ to be a scaled identity matrix for simplicity; we may also use the existing beamforming solutions, such as the ZF or MMSE beamforming in \cite{Jou10} as $\boldsymbol{A}^{0}$, for initialization. The choice of $\boldsymbol{A}^{0}$ does not significantly affect the convergence speed.
\end{remark}

\begin{remark}
The optimality of the solution given by Algorithm 1 to the original
problem in (\ref{eq.maxmin2k}) depends on the rank of the solution
matrix $\boldsymbol{X}^{\mathrm{opt}}$. If Algorithm 1 yields a
rank-one $\boldsymbol{X}^{\text{opt}}$ for (\ref{eq.relax}), then we
find the optimal $\boldsymbol{a}^{\text{opt}}$ as the (scaled)
eigenvector with respect to the only positive eigenvalue of
$\boldsymbol{X}^{\text{opt}}$, and obtain optimal beamforming matrix
$\boldsymbol{A}^{\text{opt}}$ for the original problem
(\ref{eq.maxmin2k}) by ``de-stacking'' the $MM\times 1$ vector
$\boldsymbol{a}^{\text{opt}}$ into a $M\times M$ matrix. In fact,
for the two-user case, it was shown in \cite{Zha09,Wan12} that
the problem (\ref{eq.sdp}), and consequently (\ref{eq.relax}),
always has a rank-one optimal solution
$\boldsymbol{X}^{\text{opt}}$. However, for the general $K>1$ case,
the existence of a rank-one optimal solution for (\ref{eq.sdp})
cannot be provably guaranteed; see also \cite{Tao12}. Hence, the
exact optimal solution for the original problem (\ref{eq.maxmin2k})
may not be constructed from the optimal
$\boldsymbol{X}^{\text{opt}}$ for its relaxed problem
(\ref{eq.relax}), the solution to which possibly has a rank greater
than one. Randomized rounding is a widely adopted method to obtain a
feasible rank-one approximate solution from the SDP relaxation;
specifically, a Gaussian randomized rounding strategy \cite{Ger10} can
be applied to get a vector $\boldsymbol{a}^{\text{opt}}$ from
$\boldsymbol{X}^{\text{opt}}$ to nicely approximate the solution of
the original problem (\ref{eq.maxmin2k}).

It is worth mentioning that, for the case of $K>1$, the output value
of Algorithm 1, obtained by dropping the rank constraint, is an
upper bound of the solution to the original max-min SINR problem in
(\ref{eq.maxmin2k}). This upper bound can be used as a benchmark to
assess the approximate solution obtained by randomized rounding.
\end{remark}


\subsection{Power Minimization Problem}

We next describe the SINR balancing problem in the form of power minimization. We show that, for the two alternative
forms of the SINR balancing problem, the solution to one can be obtained through solving the other.

The power minimization problem is formulated as follows:
\begin{equation}\label{eq.power}
\begin{split}
& \min_{\boldsymbol{A}} \; \sum_{i=1}^{2K} {p_i} \|\boldsymbol{A}\boldsymbol{h}_i\|^2 + \text{tr}(\boldsymbol{A} \boldsymbol{\Lambda}_R \boldsymbol{A}^H)\\
& \text{s. t.}  ~~~ \text{SINR}_i(\boldsymbol{A}) \geq \gamma_i, ~~i =1,\ldots, 2K.
\end{split}
\end{equation}
Noting $\boldsymbol{a} = \mathrm{vec}(\boldsymbol{A})$ and $\boldsymbol{X} = \boldsymbol{a}\boldsymbol{a}^H$, and dropping the rank constraint of $\boldsymbol{X}$, we can rewrite (\ref{eq.power}) as
\begin{equation}\label{eq.power33}
\begin{split}
& P_R(\lambda) = \min_{\boldsymbol{X}\succeq 0} \; \text{tr}(\boldsymbol{E}_0 \boldsymbol{X})  \\
& \text{s. t.}~~\frac{f_i(\boldsymbol{X})}{g_i(\boldsymbol{X})} \geq
\lambda\gamma_i, ~~i =1,\ldots,2K.
\end{split}
\end{equation}
Clearly, setting the parameter $\lambda$  to 1 reduces (\ref{eq.power33}) to (\ref{eq.power}).
Here, we allow $\lambda$ to be an arbitrary positive number for ease of further discussions. We note that the power
minimization in (\ref{eq.power33}) can be efficiently solved with a
single SDP \cite{Tao12}.

We next establish a close relation between the max-min SINR problem in (\ref{eq.maxmin2k}) and the power minimization problem in (\ref{eq.power33}).
We first show that (\ref{eq.power33}) can be solved via solving (\ref{eq.maxmin2k}). Let $\tilde{\lambda}^{\text{opt}}(\check{P}_R)$ denote the optimal
value of (\ref{eq.relax}) for a given power budget $\check{P}_R$. It can be shown that $\tilde{\lambda}^{\text{opt}}(\check{P}_R)$ is
a strictly increasing function of $\check{P}_R$, and the optimal
solution to (\ref{eq.power}) is the same as that to (\ref{eq.maxmin2k}) with the power
budget $P_R$ satisfying $\tilde{\lambda}^{\text{opt}}(P_R) = 1$. (See the Appendix for proof.)
As a result, the optimal solution to (\ref{eq.power}) can be
obtained by solving the equation $\tilde{\lambda}^{\text{opt}}(P_R)
= 1$, which simply requires a one-dimensional bisection search.

What remains is to show that (\ref{eq.maxmin2k}) can be solved via solving (\ref{eq.power33}).
It can be similarly shown that $P_R(\lambda)$ in (\ref{eq.power33}) is a strictly increasing function of $\lambda$. Together with the fact that, for an arbitrary $\lambda > 0$, (\ref{eq.power33}) is readily solvable using a single SDP, we conclude that (\ref{eq.maxmin2k}) is solvable by a bisection search over $\lambda$ satisfying $P_R(\lambda)=\check{P}_R$.

So far, we have shown that the power minimization and max-min SINR problems are two alternative forms of the SINR balancing
problem. This allows us to freely choose a more tractable form, i.e., a form that is more efficiently solvable, as the corner stone to pursue the optimal
beamforming designs under various important optimization criteria, as detailed in what follows.

\section{A Unified Approach via Monotonic Program}
In this section, using the max-min SINR or power minimization
solution as a corner stone, we propose a unified approach to find
the relay beamforming designs for sum rate maximization, sum MSE
minimization, and average BER minimization, etc.

\subsection{Some Useful Definitions}

We start with some commonly used terminologies in monotonic programming \cite{Tuy00}:

{\bf Definition 1 (Box):} A {\em box} $[\boldsymbol{0},
\boldsymbol{b}]$ is defined as the set of all $\boldsymbol{z}$ such
that $\boldsymbol{0} \leq \boldsymbol{z} \leq \boldsymbol{b}$.

{\bf Definition 2 (Normal):} A set ${\cal S}$ is called {\em normal}
if $\boldsymbol{z}' \leq \boldsymbol{z}$ and $\boldsymbol{z} \in
{\cal S}$ implies $\boldsymbol{z}' \in {\cal S}$.

{\bf Definition 3 (Reverse Normal):} A set ${\cal S}$ is called {\em
reverse normal} if $\boldsymbol{z}' \geq \boldsymbol{z}$ and
$\boldsymbol{z} \in {\cal S}$ implies $\boldsymbol{z}' \in {\cal
S}$.

{\bf Definition 4 (Polyblock):} For any finite vector set ${\cal
T}:=\{\boldsymbol{v}_j | j=1,\ldots, J\}$, the union of all the boxes
$[\boldsymbol{0}, \boldsymbol{v}_j]$, $\forall j$, is a {\em
polyblock} with vertex set ${\cal T}$.

{\bf Definition 5 (Proper):} A vertex $\boldsymbol{v}_j \in {\cal
T}$ is called {\em proper} if there does not exist another
$\boldsymbol{v}_{j'} \in {\cal T}$ such that $\boldsymbol{v}_{j'}
\geq \boldsymbol{v}_j$. A polyblock is fully determined by its
proper vertices.

{\bf Definition 6 (Projection):} For any $\boldsymbol{z} \in
\mathbb{R}_+^{2K} \backslash \{\boldsymbol{0}\}$ and a normal set
${\cal G}$, $\pi_{\cal G}(\boldsymbol{z})$ is a {\em projection} of
$\boldsymbol{z}$ on ${\cal G}$ if $\pi_{\cal G}(\boldsymbol{z}) =
\lambda \boldsymbol{z}$ where $\lambda = \max\{\alpha\;|\;
\alpha\boldsymbol{z} \in {\cal G}\}$; i.e., $\pi_{\cal
G}(\boldsymbol{z})$ is the unique point where the halfline from
$\boldsymbol{0}$ through $\boldsymbol{z}$ meets the upperboundary of
${\cal G}$.

\subsection{Weighted Sum-Rate Maximization}

Now consider the beamforming design for weighted sum-rate maximization. Treat the inter-user interference as noise.
For the $\text{SINR}_i(\boldsymbol{A})$ in (\ref{eq.sinr1}) and
(\ref{eq.sinr2}), we adopt a Shannon-capacity rate
formula $r_i(\boldsymbol{A}) = 0.5 \log_2(1+
\text{SINR}_i(\boldsymbol{A}))$ due to its wide applications in
communication systems. The results will be generalized to other
utility functions in the sequel. Let $w_i$ denote the priority weight
for user $i$. We aim to solve the weighted sum-rate maximization
problem formulated as
\begin{equation}\label{eq.prate}
\begin{split}
\ \max_{\boldsymbol{A}} ~~~ & \sum_{i=1}^{2K} 0.5 w_i \log_2(1+ \text{SINR}_i(\boldsymbol{A}))\\
\text{s. t.}  ~~~ & \sum_{i=1}^{2K} {p_i} \|\boldsymbol{A}\boldsymbol{h}_i\|^2 + \text{tr}(\boldsymbol{A} \boldsymbol{\Lambda}_R \boldsymbol{A}^H) \leq \check{P}_R.
\end{split}
\end{equation}

In terms of $\boldsymbol{X} = \mathrm{vec}{\mathbf{A}}(\mathrm{vec}{\mathbf{A}})^H$, we rewrite (\ref{eq.prate}) as
\begin{equation}\label{eq.prate1}
\ \max_{\boldsymbol{X}\succeq 0} ~\sum_{i=1}^{2K} 0.5 w_i \log_2(1+ \text{SINR}_i(\boldsymbol{X})), \quad
\text{s. t.}  ~~ \text{tr}(\boldsymbol{E}_0 \boldsymbol{X}) \leq \check{P}_R
\end{equation}
where $\text{SINR}_i(\boldsymbol{X}) = f_i(\boldsymbol{X}) /g_i(\boldsymbol{X})$. Note that the rank constraint of $\boldsymbol{X}$ is dropped in (\ref{eq.prate1}), and thus (\ref{eq.prate1}) is in fact a relaxation of (\ref{eq.prate}).

Define the set ${\cal X}:=\{\boldsymbol{X}\;|\;  \text{tr}(\boldsymbol{E}_0 \boldsymbol{X}) \leq \check{P}_R\}$.
Introducing an auxiliary vector $\boldsymbol{z}=[z_1, \ldots,
z_{2K}]^T$, we can reformulate (\ref{eq.prate1}) into
\begin{equation}\label{eq.prate2}
    \max_{\boldsymbol{z} \in {\cal Z}} \;\Phi(\boldsymbol{z}):=\sum_{i=1}^{2K} 0.5 w_i \log_2(z_i),
\end{equation}
where the feasible set ${\cal Z}:=\{\boldsymbol{z}\;|\; 1 \leq z_i \leq 1+\text{SINR}_i(\boldsymbol{X}), i=1,\ldots, 2K, \; \forall \boldsymbol{X} \in {\cal X}\}$.
Let $\boldsymbol{z}^{\text{opt}}$ be the optimal solution to (\ref{eq.prate2}). Then, $\boldsymbol{X}^{\text{opt}} \in {\cal X}$ satisfying $z_i^{\text{opt}} = 1+\text{SINR}_i(\boldsymbol{X}^{\text{opt}})$ for all $i$ is clearly the optimal solution to the original problem (\ref{eq.prate1}).

Now let
\begin{equation}\label{GG}
{\cal G}:= \{\boldsymbol{z}\;|\; 0 \leq z_i \leq 1+
\text{SINR}_i(\boldsymbol{X}), \forall i, \; \forall \boldsymbol{X}
\in {\cal X}\}.
\end{equation}
Also let $\boldsymbol{b}(\boldsymbol{X}):=[1+\text{SINR}_1(\boldsymbol{X}),
\ldots, 1+\text{SINR}_{2K}(\boldsymbol{X})]^T$, for any
$\boldsymbol{X} \in {\cal X}$. Then ${\cal G}=\cup_{\boldsymbol{X}
\in {\cal X}} [\boldsymbol{0}, \boldsymbol{b}(\boldsymbol{X})]$, implying that
${\cal G}$ can be represented as the union of an infinite number of normal boxes; hence, ${\cal G}$ is also normal \cite{Tuy00}.
Let $\boldsymbol{d}:=[d_1, \ldots, d_{2K}]^T$, with
\begin{equation}\label{eq.di}
    d_{2k-1} = 1+\frac{p_{2k} \check{P}_R \|\boldsymbol{h}_{2k-1}\|^2 \|\boldsymbol{h}_{2k}\|^2}{\sigma_{2k-1}^2}, \qquad
    d_{2k}  = 1+\frac{p_{2k-1} \check{P}_R \|\boldsymbol{h}_{2k-1}\|^2 \|\boldsymbol{h}_{2k}\|^2}{\sigma_{2k}^2}.
\end{equation}
It clearly holds: $1+\text{SINR}_i(\boldsymbol{X}) \leq d_i$,
$\forall i$, $\forall \boldsymbol{X} \in {\cal X}$. Therefore,
${\cal G} \subset [\boldsymbol{0}, \boldsymbol{d}]$ is a compact
normal set with nonempty interior.
Further define ${\cal H}:=\{\boldsymbol{z}\;|\; z_i \geq 1, \forall i\}$. Clearly, ${\cal H}$
is a reverse normal set. Then (\ref{eq.prate2}) can be written in the form of a
standard MP \cite{Tuy00} as
\begin{equation}\label{eq.mp}
    \max_{\boldsymbol{z}} \;\Phi(\boldsymbol{z}), ~~~ \text{s. t.}  ~~ \boldsymbol{z} \in  {\cal G} \cap {\cal H}.
\end{equation}



For the MP (\ref{eq.mp}), a polyblock outer approximation method can
be employed to efficiently find its global optimal solution \cite{Tuy00}. Specifically, we target at constructing a
nested sequence of polyblocks ${\cal P}_n$, $n=1,2,\ldots$,
approximating ${\cal G} \cap {\cal H}$: ${\cal P}_1 \supset {\cal
P}_2 \supset \cdots \supset {\cal G} \cap {\cal H}$ in such a way
that $\max_{\boldsymbol{z} \in {\cal P}_n} \; \Phi(\boldsymbol{z}) ~
\searrow ~ \max_{\boldsymbol{z} \in {\cal G} \cap {\cal H}} \;
\Phi(\boldsymbol{z})$. Denote the maximizer at iteration $n$ as
\begin{equation}\label{eq.zn}
    \boldsymbol{z}^n = \arg \max_{\boldsymbol{z} \in {\cal T}_n} \; \Phi(\boldsymbol{z}),
\end{equation}
where ${\cal T}_n$ is the (finite) proper vertex set of ${\cal P}_n$. Note that $\boldsymbol{z}^n$ can be obtained by exhaustively searching over the finite set ${\cal T}_n$. If $\boldsymbol{z}^n \in {\cal G} \cap {\cal H}$, then it solves the MP in (\ref{eq.mp}). Otherwise, we find the next polyblock ${\cal P}_{n+1}$ contained in ${\cal P}_n$ but still containing ${\cal G} \cap {\cal H}$, and continue the process.

We next find ${\cal P}_{n+1}$ from ${\cal P}_n$. Let $\boldsymbol{y}^n$ be the projection of $\boldsymbol{z}^n$ on $\cal G$, i.e., $\boldsymbol{y}^n =
\pi_{\cal G}(\boldsymbol{z}^n)$, and denote
\begin{equation}\label{eq.znk}
    \boldsymbol{z}^{n}(i) = \boldsymbol{z}^n - (z^n_i - y^n_i) \boldsymbol{e}_i, \quad i = 1,\ldots
    2K,
\end{equation}
where $\boldsymbol{e}_i$ is a unit vector with the only non-zero (i.e., ``1'') in the $i$-th entry.
Note that $\boldsymbol{z}^{n}(i)$ is obtained by replacing the $i$-th entry of $\boldsymbol{z}^{n}$ by
$y^n_i$. Clearly, $\boldsymbol{y}^n \leq \boldsymbol{z}^{n}(i) \leq
\boldsymbol{z}^n$. Let ${\cal T}_{n+1}$ be the set obtained from
${\cal T}_n$ by replacing the vertex $\boldsymbol{z}^n$ with $2K$
new vertices $\boldsymbol{z}^{n}(i)$ and then remove the improper
vertices; i.e., ${\cal T}_{n+1}=({\cal T}_n \backslash
\{\boldsymbol{z}^n\})\cup
\{\boldsymbol{z}^{n}(i)\;|\;\boldsymbol{z}^{n}(i)~\text{is proper}\}$.
Since $\boldsymbol{z}^{\text{opt}} \in {\cal H}$, we can further
reduce the vertex set ${\cal T}_{n+1} = {\cal T}_{n+1} \cap {\cal
H}$. From \cite[Proposition 17]{Tuy00}, we immediately have
\begin{lemma}
The polyblock ${\cal P}_{n+1}$ with vertex set ${\cal T}_{n+1}$ satisfies $({\cal G} \cap {\cal H}) \subset {\cal P}_{n+1} \subset {\cal P}_n$.
\end{lemma}

Lemma 2 guarantees the validity of the above constructed ${\cal P}_{n+1}$ to continue the polyblock outer approximation
process. A key step in the above construction of ${\cal P}_{n+1}$ is to find the projection
$\boldsymbol{y}^n = \pi_{\cal
G}(\boldsymbol{z}^n)=\lambda^n\boldsymbol{z}^n$, which can be
determined by solving
\begin{align}
    \nonumber \lambda^n & = \max\{\alpha\;|\;\alpha\boldsymbol{z}^n \in {\cal G}\} \\
    \nonumber & = \max\{\alpha\;|\; \alpha \leq \min_{i=1,\ldots,2K} \frac{1+\text{SINR}_i(\boldsymbol{X})}{z^n_i}, \; \forall \boldsymbol{X} \in {\cal X}\} \\
     & = \max_{\boldsymbol{X} \in {\cal X}} \; \min_{i=1,\ldots,2K} \frac{1+\text{SINR}_i(\boldsymbol{X})}{z^n_i},
\end{align}
where the second step utilizes the definition of $\cal G$ in (\ref{GG}).
The above is an extended max-min SINR balancing problem written as
\begin{equation}\label{eq.pt}
\begin{split}
\lambda^n  = &\max_{\boldsymbol{X}}\; \min_{i=1,\ldots,2K}  \frac{1+\text{SINR}_i(\boldsymbol{X})}{z^n_i}\\
& \text{s. t.}  ~~~ \boldsymbol{X}\succeq 0,\quad \text{tr}(\boldsymbol{E}_0 \boldsymbol{X}) \leq \check{P}_R.
\end{split}
\end{equation}
This problem can be solved using the Dinkelbach-type
Algorithm 1 with minor modifications. Use the definitions in Section II (such as
$\boldsymbol{\Phi}$, $\boldsymbol{q}_{ji}$, $\boldsymbol{B}_i$, and $g_i(\boldsymbol{X})$), except that $f_i(\boldsymbol{X})$ is redefined as
$f_i(\boldsymbol{X}):= \text{tr}(\boldsymbol{E}_i^{(1)} \boldsymbol{X}) + \text{tr}(\boldsymbol{E}_i^{(2)}\boldsymbol{X}) + \sigma_i^2$.
Then the solution of (\ref{eq.pt}) can be obtained by solving a series of
(\ref{eq.maxminX2k}).




We are now ready to implement polyblock outer approximation method for (\ref{eq.prate}). 
For a given accuracy tolerance level $\epsilon>0$, we say that a
feasible $\bar{\boldsymbol{z}}$ is an $\epsilon$-optimal solution if
$(1+\epsilon) \Phi(\bar{\boldsymbol{z}}) \geq
\Phi(\boldsymbol{z}^{\text{opt}})$. The following algorithm is
proposed to find an $\epsilon$-optimal solution for
(\ref{eq.prate1}).

\begin{quote}
\vspace{0.05 in}\hrule height0.1pt depth0.3pt \vspace{0.05 in}
\begin{algorithm}\label{algorithm.2}
{\it for weighted sum-rate maximization}

{\bf Initialize}: select an accuracy level $\epsilon>0$, let $n=0$,
${\cal T}_0 = \{\boldsymbol{d}\}$, and $\text{CBV} = -\infty$.

{\bf Repeat}: \\
    1). let $\boldsymbol{z}^n = \arg \max_{\boldsymbol{z} \in {\cal T}_n} \; \Phi(\boldsymbol{z})$,
    For $\boldsymbol{z}^n$, use Algorithm 1 to solve (\ref{eq.pt}) to obtain $\lambda^n$, and the corresponding $\boldsymbol{X}^{\text{opt}}$, as well as $\boldsymbol{y}^n = \lambda^n\boldsymbol{z}^n$.\\
    2). If $\boldsymbol{y}^n \in {\cal H}$ and $\Phi(\boldsymbol{y}^n) > \text{CBV}$, then $\text{CBV}=\Phi(\boldsymbol{\boldsymbol{y}^n})$, $\bar{\boldsymbol{z}}=\boldsymbol{y}^n$ and $\bar{\boldsymbol{X}}=\boldsymbol{X}^{\text{opt}}$. \\
    3). Let $\boldsymbol{z}^{n}(i) = \boldsymbol{z}^n - (z^n_i - y^n_i) \boldsymbol{e}_i$, $\forall i $, and
    ${\cal T}_{n+1}=[({\cal T}_n \backslash \{\boldsymbol{z}^n\})\cup \{\text{proper}~ \boldsymbol{z}^{n}(i)\}] \cap {\cal H}$. \\
    4). Further remove from ${\cal T}_{n+1}$ any $\boldsymbol{v}_j \in {\cal T}_{n+1}$ satisfying $\Phi(\boldsymbol{v}_j) \leq \text{CBV}(1 + \epsilon)$.\\
    5). Set $n=n+1$.

{\bf until} ${\cal T}_n = \phi$.

{\bf Output}: $\bar{\boldsymbol{z}}$ as the $\epsilon$-optimal solution for (\ref{eq.prate2}) and $\bar{\boldsymbol{X}}$ the solution for (\ref{eq.prate1}).
\end{algorithm}
\vspace{0.05 in} \hrule height0.1pt depth0.3pt \vspace{0.1 in}
\end{quote}

Per iteration $n$ of Algorithm 2, we have $\boldsymbol{y}^n =
\pi_{\cal G}(\boldsymbol{z}^n)  \in {\cal G}$. If $\boldsymbol{y}^n
\in {\cal H}$ is also true, we obtain a feasible point $\boldsymbol{y}^n \in
{\cal G} \cap {\cal H}$. In this case, we update $\text{CBV} =
\max\{\text{CBV}, \Phi(\boldsymbol{y}^n)\}$. This implies that $\text{CBV}$ is
the current best value so far, and the corresponding
$\bar{\boldsymbol{z}}=\arg \max_{\{\boldsymbol{y}^m\; |\;
\boldsymbol{y}^m \in {\cal H}, m \leq n\}} \Phi(\boldsymbol{y}^m)$
is the current best solution for (\ref{eq.prate2}). Observe that for
any $\boldsymbol{v}_j \in {\cal T}_{n+1}$ satisfying
$\Phi(\boldsymbol{v}_j) \leq \text{CBV}(1 + \epsilon)$, we have
$(1+\epsilon)\text{CBV} \geq \Phi(\boldsymbol{y})$, $\forall
\boldsymbol{y} \in [\boldsymbol{0}, \boldsymbol{v}_j]$, due to
monotonicity of $\Phi$. Hence, $\boldsymbol{v}_j$ can be removed
from ${\cal T}_{n+1}$ for further consideration since
$\bar{\boldsymbol{z}}$ will be the desired $\epsilon$-optimal
solution if $\boldsymbol{z}^{\text{opt}} \in [\boldsymbol{0},
\boldsymbol{v}_j]$.

\begin{remark}
We remark that Algorithm \ref{algorithm.2} yields the $\epsilon$-optimal solution to (\ref{eq.prate}) for the case of $K=1$. However, for the general case of $K>1$, the output value of Algorithm \ref{algorithm.2}, obtained by dropping the rank constraint, only provides an upper bound of the maximum weighted sum-rate of (\ref{eq.prate}). Again, randomized rounding is used to obtain a good approximate solution to (\ref{eq.prate}).
\end{remark}

\begin{figure}
\centering \epsfig{file=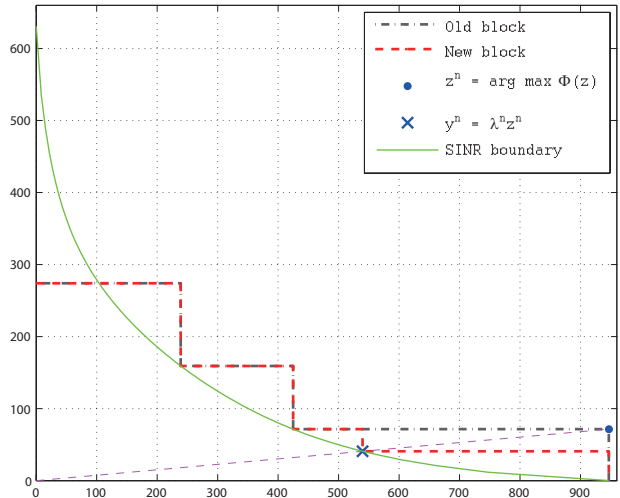, width=0.55\textwidth} \caption{
The polyblock outer approximation procedure.} \label{F:fig2}
\vspace{-0.3cm}
\end{figure}

An illustration of Algorithm \ref{algorithm.2} for $K = 1$ is given in Fig.
\ref{F:fig2}. With a vertex set ${\cal T}_{n}$, the upperboundary of
polyblock ${\cal P}_n$ is depicted by the black dotted-dashed line.
Among the three entries of ${\cal T}_n$, the third one is the
maximizer: $\boldsymbol{z}^n = \arg \max_{\boldsymbol{z} \in {\cal
T}_n} \; \Phi(\boldsymbol{z})$, which is marked with a blue dot.
After finding its projection $\boldsymbol{y}^n$ (marked with a blue
cross) on the achievable SINR boundary, two new vertices
$\boldsymbol{z}^{n,1}$ and $\boldsymbol{z}^{n,2}$ are then obtained
through (\ref{eq.znk}). By replacing $\boldsymbol{z}^n$ with these
two vertices, we determine the new polyblock ${\cal P}_{n+1}$ with
its upperboundary given by the red dashed line.

Similar polyblock outer approximation approaches have been adopted
to solve the linear fractional programming and non-convex wireless
power control problems in \cite{Phu03, Qia09}. A key requirement for
provable convergence of Algorithm \ref{algorithm.2} is that $\boldsymbol{z}$ is
lower bounded by a strictly positive vector. Since $\boldsymbol{z}
\geq \boldsymbol{1} > \boldsymbol{0}$ in (\ref{eq.prate2}), it
readily follows from \cite[Theorem 1]{Tuy00} that
\begin{proposition}
Algorithm 2 globally converges to an $\epsilon$-optimal solution for
(\ref{eq.prate2}) and (\ref{eq.prate1}).
\end{proposition}

The proposed Algorithm 2 can yield optimal TWR beamforming solution
for the relaxed weighted throughput maximization (\ref{eq.prate1})
with guaranteed convergence and global optimality. For the
two-user case, the algorithm can also yield the globally optimal
solution for the original problem (\ref{eq.prate}); for the general $K$-pair
case, it can provide a good approximate solution for
(\ref{eq.prate}). Hence, the proposed approach provides a good
benchmark for all the beamforming (or precoding) schemes that are
designed to maximize the user rates in AF-based TWR.

Note that the outer polyblock approximation is in fact a
branch-and-bound method. For coordinated beamforming designs in
multicell networks, a branch-reduce-and-bound (BRB) algorithm was
proposed. It was shown that this BRB algorithm can have faster convergence for weighted sum-rate maximization problems, whereas the polyblock approximation has faster convergence for many other utility functions \cite{Bjo12}. The key in the BRB algorithm is again
finding the projection of an outer vertex on the upperboundary of the
achievable SINR region. Using the max-min SINR solution for
(\ref{eq.maxmin2k}), a BRB algorithm similar to Algorithm 2 can be
also developed to find the optimal TWR beamforming design for the
weighted throughput maximization (\ref{eq.prate}), probably with
a faster convergence speed.

\subsection{General Design Criteria}

The proposed MP approach only relies on the monotonicity of
the objective function and the normality of the feasible set. Thus, it can
apply to beamforming designs under more general criteria. Consider
maximizing a general increasing function $F_i$ of SINRs
\begin{equation}\label{eq.p0f}
\begin{split}
\ \max_{\boldsymbol{A}} ~~~&\sum_{i=1}^{2K} F_i(\text{SINR}_i(\boldsymbol{A}))\\
\text{s. t.}  ~~~ & \sum_{i=1}^{2K} {p_i} \|\boldsymbol{A}\boldsymbol{h}_i\|^2 + \text{tr}(\boldsymbol{A} \boldsymbol{\Lambda}_R \boldsymbol{A}^H) \leq \check{P}_R.
\end{split}
\end{equation}

The function $F_i$ can be a specific rate function (different from the Shannon capacity formula) $r_i(\text{SINR}_i(\boldsymbol{A}))$ for practical modulation and coding schemes. Maximization of the utility of user rates has gained a growing interest in the communication and networking context, where different types of utility functions are proposed to trade off the throughput and fairness, or to capture the ``happiness'' of the user links \cite{Luo08}. The function $F_i$ here can also be the composition of an increasing (not necessarily concave) utility function with that particular rate function $U_i(r_i(\text{SINR}_i(\boldsymbol{A})))$.

In addition, the formulation (\ref{eq.p0f}) includes the following two important cases:
\begin{enumerate}
\item {MSE minimization}: Assume that all the user receivers use the linear-minimum-mean-square-error (LMMSE) filters for estimating the received symbols. The weighted sum-MSEs at the output of the LMMSE receivers is given by \cite{Vis99}:
    \[
    \sum_{i=1}^{2K} w_i \text{MSE}_i = \sum_{i=1}^{2K} \frac{w_i}{1+\text{SINR}_i}.
    \]
    With $F_i(\text{SINR}_i(\boldsymbol{A})):= -\frac{w_i}{1+\text{SINR}_i(\boldsymbol{A})}$, (\ref{eq.p0f}) specializes to weighted sum-MSE minimization.

\item{SER or BER minimization}: Using a Q-function: $Q(x):=\frac{1}{\sqrt{2\pi}} \int_{x}^{\infty} \exp(-\frac{u^2}{2}) du$, the SER and BER of practical modulation schemes can be calculated or approximated in closed-form \cite{Gold05}. Clearly all these SER or BER functions, say $\varepsilon_i(\text{SINR}_i)$, are strictly decreasing in SINR.  With $F_i(\text{SINR}_i(\boldsymbol{A})):= -w_i\varepsilon_i(\text{SINR}_i(\boldsymbol{A}))$, the problem (\ref{eq.p0f}) specializes to weighted sum-SER (or BER) minimization.

\end{enumerate}

It is clear that (\ref{eq.p0f}) also carries over to minimization of increasing (not necessarily convex) cost functions of MSE, SER or BER.

%

For all these $F_i(\text{SINR}_i(\boldsymbol{A}))$ functions, we can
redefine $\Phi(\boldsymbol{z}):=\sum_{i=1}^{2K} F_i(z_i-1)$, and
consider
\begin{equation}
    \max_{\boldsymbol{z} \in {\cal Z}} \;\Phi(\boldsymbol{z}):=\sum_{i=1}^{2K} F_i(z_i-1).
\end{equation}
Algorithm 2 can be used to approximately solve
this MP, and, subsequently, provide the solution for (\ref{eq.p0f}).
It provides a benchmark for the beamforming designs in AF-based TWR
under many important criteria.

\section{Collaborative TWR Beamforming}

\subsection{Collaborative TWR Model}

The proposed unified framework also applies to collaborative TWR where a cluster of $M$ single-antenna relay nodes $\{R_m \mid
m=1, \ldots, M\}$ cooperatively assist the bidirectional
communications between multiple users. Such a collaborative TWR
scheme was previously considered in \cite{ZHKV12},\cite{CGHMW11} and
\cite{Wang11}, where the beamforming coefficients for the
relays are designed under a total relay power constraint, i.e., the relays share
a total power budget. This total relay power constraint is usually
not realistic in practical scenarios. Therefore, we consider
collaborative beamforming design with individual relay power
constraints.

The system model for collaborative TWR can be viewed as a special
case of the TWR model described in Section II. The only difference
is that in collaborative TWR, the signals received by different
antennas at relays cannot be jointly processed. Assume that the $(2k-1)$th user and the $(2k)$th user
communicate with each other, $k=1, \ldots, K$, and that data exchange consists of two phases. In the first phase, each
user transmits its signal $s_i(t)$ to the relays, and the received
signal $y_{R_m}(t)$ at the relay $R_m$ is
\begin{equation}
    y_{R_m}(t) = \sum_{i=1}^{2K} h_{i,m} \sqrt{p_i} s_i(t) + n_{R_m}(t),
\end{equation}
where $h_{i,m}$ denotes the channel coefficient from user $i$ to
relay $R_m$, and $z_{R_m}(t) \thicksim {\cal CN}(0,\sigma_{R_m}^2)$
denotes the additive noise at relay $R_m$. Let
$\boldsymbol{h}_i:=[h_{i,1},\ldots, h_{i,M}]^T$,
$\boldsymbol{y}_R(t):=[y_{R_1}(t),\ldots, y_{R_M}(t)]^T$, and
$\boldsymbol{z}_R(t):=[z_{R_1}(t),\ldots, z_{R_M}(t)]^T$. Then the
received signal vector $\boldsymbol{y}_R(t)$ at all relays is again
given by (\ref{eq.yR}).

Upon receiving $y_{R_m}(t)$, the relay collaboratively amplifies and
forwards its signal $x_{R_m}(t) = \tilde{a}_m y_{R_m}(t)$ to all
users in the next phase. Let $\boldsymbol{\tilde{a}}:=[\tilde{a}_1,
\ldots, \tilde{a}_M]$ collect the (complex) AF gains for all relays.
The signal vector $\boldsymbol{x}_R(t):=[x_{R_1}(t),\ldots,
x_{R_M}(t)]^T$ can be written as $   \boldsymbol{x}_R(t) =
\boldsymbol{\tilde{A}}\boldsymbol{y}_R(t)$, where
$\boldsymbol{\tilde{A}}:=\text{diag}(\boldsymbol{\tilde{a}})$.
Different from the TWR model with a multi-antenna relay in Section II, the
beamforming matrix for collaborative TWR is restricted to be
diagonal. The transmit power of the relay $R_m$ is given by
\begin{equation}
    p_{R_m}(\boldsymbol{\tilde{a}}) = \sum_{i=1}^{2K} p_i \|\tilde{a}_m \boldsymbol{h}_{i,m}\|^2 + \sigma_{R_m}^2 |\tilde{a}_m|^2.
\end{equation}

Assuming channel reciprocity, the received signal at user
$i=1,\ldots, 2K$, is then given by
\begin{equation}\label{eq.yic}
    y_i(t) = \boldsymbol{h}_i^T \boldsymbol{\tilde{A}} \sum_{j=1}^{2K} \boldsymbol{h}_j \sqrt{p_j} s_j(t) + \boldsymbol{h}_i^T \boldsymbol{\tilde{A}} \boldsymbol{n}_R(t) + n_i(t)
\end{equation}
where the noise $n_i(t) \thicksim {\cal CN}(0, \sigma_i^2)$. Clearly, (\ref{eq.yic}) is equivalent to (\ref{eq.yi}) by replacing $\boldsymbol{\tilde{A}}$
with $\boldsymbol{A}$. Therefore, after removing the self-interference, the SINR at the $(2k-1)$th user and at the $(2k)$th user are respectively given
by (\ref{eq.sinr1}) and (\ref{eq.sinr2}) (with $\boldsymbol{A}$ replaced by $\boldsymbol{\tilde{A}}$).

\subsection{Algorithm Design}

Based on these SINRs, the max-min SINR problem for collaborative TWR can be formulated as
\begin{equation}\label{eq.maxminc}
\begin{split}
\lambda^{\text{opt}} = & \max_{\boldsymbol{\tilde{A}}} \min_{i=1,\ldots, 2K} \frac{\text{SINR}_i(\boldsymbol{\tilde{A}})}{\gamma_i}\\
\text{s. t.}  ~~~& \sum_{i=1}^{2K} p_i \|\tilde{a}_m
\boldsymbol{h}_{i,m}\|^2 + \sigma_{R_m}^2 |\tilde{a}_m|^2 \leq
\check{P}_{R_m}, m=1, \ldots, M.
\end{split}
\end{equation}

Problem (\ref{eq.maxminc}) is similar to (\ref{eq.maxmin2k}) except
that $\boldsymbol{\tilde{A}}$ in (\ref{eq.maxminc}) is constrained
to be diagonal and there are $M$ transmit power constraints. Thus,
(\ref{eq.maxminc}) can be solved in a similar way as
(\ref{eq.maxmin2k}) is. Let $\boldsymbol{X} =
\boldsymbol{\tilde{a}}\boldsymbol{\tilde{a}}^H$, $ \theta_m :=
\sum_{i=1}^{2K} p_i \|\boldsymbol{h}_{i,m}\|^2 + \sigma_{R_m}^2$,
$\boldsymbol{\Phi}_m := [ \boldsymbol{0}_{1 \times (m-1)}, \theta_m,
\boldsymbol{0}_{1 \times (M-m)}]$, and $\boldsymbol{E}_{0,m} :=
\boldsymbol{\Phi}_m^H \boldsymbol{\Phi}_m $. Then the transmit power
constraint of relay $R_m$ can be expressed as
$\text{tr}(\boldsymbol{E}_{0,m} \boldsymbol{X}) \leq
\check{P}_{R_m}$. Upon defining $f_i(\boldsymbol{X})$ and $g_i(\boldsymbol{X})$ as with (\ref{eq.fgi}), the problem (\ref{eq.maxminc}) can be relaxed to a max-min fractional program similar to (\ref{eq.relax}). Consequently, it can be efficiently solved by the Dinkelbach-type Algorithm 1 with minor modifications.

Using the max-min SINR solution as the corner stone, the beamforming
designs for the collaborative TWR under the various criteria considered in Section IV can be done with minor modifications of
Algorithm 2. For example, the weighted sum-rate maximization problem for
collaborative TWR is the same as (\ref{eq.mp}) except that the set
${\cal X}$ is now given by ${\cal X}:=\{\boldsymbol{X}\;|\;
\text{tr}(\boldsymbol{E}_{0,m} \boldsymbol{X}) \leq \check{P}_{R_m},
m=1,\ldots,M \}$. It is clear that the corresponding set ${\cal G}$
for collaborative TWR is still normal. Hence, the optimization
problem can be still formulated as an MP, and the optimal beamforming matrix can be obtained using the polyblock outer approximation method in Algorithm 2.

\section{MIMO TWR Beamforming}

\subsection{MIMO TWR Model}

The performance of TWR can be enhanced when both the
relay and the users are equipped with multiple antennas
\cite{Wan12}. In what follows, we consider the joint optimization of
users' transmit and receive beamforming vectors and the relay's
beamforming matrix.

Let $M_i$ denote the number of antennas at user $i=1,\ldots, 2K$,
and $s_i(t)$ denote the data signal. In the first phase, user $i$
performs transmit beamforming with vector $\boldsymbol{u}_i \in
\mathbb{C}^{M_i \times 1}$ as $\boldsymbol{x}_i(t) =
\boldsymbol{u}_i s_i(t)$, where $||\boldsymbol{u}_i||^2 \le p_i$,
and $p_i$ is the transmit power budget of user $i$. The received
signal at the relay is
\begin{equation}\label{eq.yRmimo}
    \boldsymbol{y}_R(t) = \sum_{i=1}^{2K} \boldsymbol{H}_i \boldsymbol{x}_i(t) + \boldsymbol{n}_R(t),
\end{equation}
where $\boldsymbol{H}_i \in \mathbb{C}^{M \times M_i}$ is the
channel matrix from user $i$ to the relay.

In the second phase, the relay amplifies and forwards the signal
$\boldsymbol{x}_R(t)=\boldsymbol{A}\boldsymbol{y}_R(t)$ to both
users. The transmit power at the relay is given by
\begin{equation}\label{eq.pRmimo}
p_R(\boldsymbol{A}) = \sum_{i=1}^{2K} \text{tr}(\boldsymbol{A}
\boldsymbol{H}_i \boldsymbol{u}_i \boldsymbol{u}_i^H
\boldsymbol{H}_i^H \boldsymbol{A}^H) + \text{tr}(\boldsymbol{A}
\boldsymbol{\Lambda}_R \boldsymbol{A}^H).
\end{equation}
The received signal at user $i$ is given by
\begin{equation}\label{eq.yimimo}
 \boldsymbol{y}_i(t) =  \boldsymbol{H}_i^T \boldsymbol{A}  \sum_{j=1}^{2K} \boldsymbol{H}_j  \boldsymbol{x}_j(t) + \boldsymbol{H}_i^T \boldsymbol{A} \boldsymbol{n}_R(t) +
 \boldsymbol{n}_i(t),
\end{equation}
where $\boldsymbol{n}_i(t) \thicksim {\cal
CN}(\boldsymbol{0},\boldsymbol{\Lambda}_i)$ is the additive noise at
user $i$.

The user $i$ first combines its received signal with a vector
$\boldsymbol{v}_i \in \mathbb{C}^{M_i \times 1} $ to obtain $
\boldsymbol{y}'_i(t) = \boldsymbol{v}_i^H \boldsymbol{y}_i(t)$,
which can be expressed as
\begin{equation} \label{eq.yi2}
 \boldsymbol{y}'_i(t) =  \boldsymbol{v}_i^H [\boldsymbol{H}_i^T \boldsymbol{A} \sum_{j=1}^{2K}  \boldsymbol{H}_j  \boldsymbol{u}_j s_j(t) + \boldsymbol{H}_i^T \boldsymbol{A} \boldsymbol{n}_R(t) + \boldsymbol{n}_i(t)].
 \end{equation}
Clearly, the output SINR of each user depends on the relay precoding
matrix $\boldsymbol{A}$, the users' transmit precoding vectors, and
the receive combining vectors. The SINR at the user $i$ is
\begin{equation}\label{eq.sinr2kmimo}
   \text{SINR}_{i}(\boldsymbol{A},\{\boldsymbol{u}_i\},\{\boldsymbol{v}_i\})= \frac{ |\boldsymbol{v}_i^H \boldsymbol{H}_i^T \boldsymbol{A} \boldsymbol{H}_{\pi(i)} \boldsymbol{u}_{\pi(i)}|^2} {\sum_{j \neq i, \pi(i)} |\boldsymbol{v}_i^H \boldsymbol{H}_i^T \boldsymbol{A} \boldsymbol{H}_j \boldsymbol{u}_j|^2 +  \| \boldsymbol{\Lambda_{R}}^{1/2} \boldsymbol{A}^H \boldsymbol{H}_i^* \boldsymbol{v}_i \|^2 + \| \boldsymbol{\Lambda_{i}}^{1/2} \boldsymbol{v}_i \|^2
   },
\end{equation}
where $\pi(i)$ denotes the partner of user $i$, i.e., $\pi(2k-1) =
2k$ and $\pi(2k) = 2k-1, \forall k$.

\subsection{Algorithm Design}
The max-min SINR problem of the considered multi-pair MIMO TWR can
be formulated as
\begin{equation}\label{eq.maxmin2kmimo}
\begin{split}
\lambda^{\text{opt}} = & \max_{\boldsymbol{A},\{\boldsymbol{u}_i\},\{\boldsymbol{v}_i\}} \ \ \min_{i=1, \ldots, 2K} \frac{\text{SINR}_i(\boldsymbol{A},\{\boldsymbol{u}_i\},\{\boldsymbol{v}_i\})}{\gamma_i}\\
\text{s. t.}  ~~~& p_R(\boldsymbol{A}) = \sum_{i=1}^{2K}
\text{tr}(\boldsymbol{A} \boldsymbol{H}_i \boldsymbol{u}_i
\boldsymbol{u}_i^H \boldsymbol{H}_i^H \boldsymbol{A}^H) +
\text{tr}(\boldsymbol{A}
\boldsymbol{\Lambda}_R \boldsymbol{A}^H) \leq \check{P}_R,   \\
~~~&  \|\boldsymbol{u}_i \|^2 \le p_i , i = 1, \ldots, 2K.
\end{split}
\end{equation}
This optimization problem is in general difficult to solve. We
next propose an iterative algorithm to optimize $\boldsymbol{A}$,
$\{\boldsymbol{u}_i\}$, and $\{\boldsymbol{v}_i\}$ in an alternating
fashion.

\subsubsection{User Receive Combining}
Given the relay beamforming matrix $\boldsymbol{A}$ and users'
transmit precoding vectors $\boldsymbol{u}_i, i=1,\ldots,2K$, the
well-known MMSE combining can be employed at user $i$ to detect the
transmit signal from its partner user. Let
$\boldsymbol{\alpha}_{i,j}:=\boldsymbol{H}_i^T \boldsymbol{A}
\boldsymbol{H}_j \boldsymbol{u}_j \in \mathbb{C}^{M_i \times 1}$,
and $ \boldsymbol{R}_i := \sum_{j \neq i} \boldsymbol{\alpha}_{i,j}
\boldsymbol{\alpha}_{i,j}^H + \boldsymbol{H}_i^T \boldsymbol{A}
\boldsymbol{\Lambda_R} \boldsymbol{A}^H \boldsymbol{H}_i^* +
\boldsymbol{\Lambda}_i$. Then the combining vector
$\boldsymbol{v}_i$ is given by
\begin{equation}\label{eq.optvi}
\boldsymbol{v}_i =   \boldsymbol{R}_i ^{-1}
\boldsymbol{\alpha}_{i,\pi(i)}.
\end{equation}

\subsubsection{Optimal Relay Precoding}
Now consider the relay beamforming design with fixed transmit and
receive beamforming vectors at the users. Let $\boldsymbol{h}_i :=
\boldsymbol{H}_i \boldsymbol{u}_i \in \mathbb{C}^{M \times 1}$,
$\boldsymbol{g}_i := \boldsymbol{H}_i^* \boldsymbol{v}_i \in
\mathbb{C}^{M \times 1}$. The max-min optimization problem in
(\ref{eq.maxmin2kmimo}) becomes
\begin{equation}\label{eq.maxminRelaymimo}
\begin{split}
\lambda_{\boldsymbol{A}}^{\text{opt}} = & \max_{\boldsymbol{A}} \min_{i=1, \ldots, 2K}  \frac{ |\boldsymbol{g}_i^H \boldsymbol{A} \boldsymbol{h}_{\pi(i)}|^2} { \gamma_i (\sum_{j \neq i, \pi(i)} | \boldsymbol{g}_i^H \boldsymbol{A} \boldsymbol{h}_j|^2 + \|\boldsymbol{\Lambda}_R^{1/2} \boldsymbol{A}^H \boldsymbol{g}_i \|^2 + \|\boldsymbol{\Lambda}_i^{1/2} \boldsymbol{v}_i \|^2)} \\
\text{s. t.}  ~~~&  \sum_{i=1}^{2K} \| \boldsymbol{A}
\boldsymbol{h}_i \|^2 +   \text{tr}(\boldsymbol{A} \boldsymbol{\Lambda}_R \boldsymbol{A}^H) \leq \check{P}_R.  \\
\end{split}
\end{equation}
This problem has almost the same form with (\ref{eq.maxmin2k}); hence, it can be
efficiently solved by Algorithm 1.

\subsubsection{Optimal Transmit Precoding}
The users' transmit precoding vectors $\boldsymbol{u}_i,
i=1,\ldots,2K$, are also designed to maximize the minimum SINR, and
the optimization problem can be formulated as
\begin{equation}\label{eq.maxminui}
\begin{split}
 & \lambda_{\boldsymbol{u}}^{\text{opt}} = \max_{\{\boldsymbol{u}_j\}_{j=1}^{2K}} \min_{i=1, \ldots, 2K} \frac{\text{SINR}_i(\boldsymbol{u})}{\gamma_i}\\
 ~~~& \text{s. t.} ~~  \|\boldsymbol{u}_i\|^2 \leq p_i,i=1,\ldots,2K.
\end{split}
\end{equation}
Let $\boldsymbol{\beta} _{i,j} :=   \boldsymbol{H}_j ^H
\boldsymbol{A}^H  \boldsymbol{H}_i^* \boldsymbol{v}_i$, and $d_i :=
\|\boldsymbol{\Lambda}_R^{1/2} \boldsymbol{A}^H \boldsymbol{g}_i
\|^2 + \|\boldsymbol{\Lambda}_i^{1/2} \boldsymbol{v}_i \|^2$.
Define: $\boldsymbol{E}_{i,j} := \boldsymbol{\beta} _{i,j}
\boldsymbol{\beta} _{i,j}^H $, and $\boldsymbol{X}_i =
\boldsymbol{u}_i \boldsymbol{u}_i^H$. The SINR of user $i$ can be
expressed as
\begin{equation}
\text{SINR}_i(\boldsymbol{u}) = \frac{ \text{tr} (
\boldsymbol{E}_{i,\pi(i)} \boldsymbol{X}_{\pi(i)} )} {\sum_{j \neq
i, \pi(i)} \text{tr} ( \boldsymbol{E}_{i,j} \boldsymbol{X}_j)  +
d_i}.
\end{equation}
Using $\boldsymbol{X}_i$, $i=1,\ldots,2K$, as the optimization
variables and dropping the constraint of
$\text{rank}(\boldsymbol{X}_i)=1$, $i=1,\ldots,2K$, the problem
(\ref{eq.maxminui}) becomes a max-min fractional program
\begin{equation}\label{eq.relaxui}
\begin{split}
 \lambda_{\boldsymbol{u}}^{\text{opt}} = & \max_{\{\boldsymbol{X}_j\}_{j=1}^{2K}} \min_{i=1, \ldots, 2K} \frac{ \text{tr} ( \boldsymbol{E}_{i,\pi(i)} \boldsymbol{X}_{\pi(i)} )} { \gamma_i (\sum_{j \neq i, \pi(i)} \text{tr} ( \boldsymbol{E}_{i,j} \boldsymbol{X}_j)  + d_i)} \\
\text{s. t.}  ~~~& \boldsymbol{X}_i \succeq 0, \quad i=1,\ldots,2K, \\
~~~&  \text{tr}(\boldsymbol{E}_i \boldsymbol{X}_i) \leq p_i,
i=1,\ldots,2K.
\end{split}
\end{equation}
Again, the problem is similar to (\ref{eq.relax}); it can be efficiently solved using the
Dinkelbach-type Algorithm 1 with minor modifications.

\subsubsection{Overall Iterative Algorithm}

We are now ready to present the overall iterative algorithm to alternatingly optimize the users' transmit precoding vectors, the relay's
beamforming matrix, and the users' receive combining vectors.

\begin{quote}
\vspace{0.05 in}\hrule height0.1pt depth0.3pt \vspace{0.05 in}
\begin{algorithm}
{\it Iterative optimization for multi-pair MIMO TWR}

{\bf Initialize}:  $\boldsymbol{u}_i^{0}, \boldsymbol{A}^{0}$, and
$\boldsymbol{v}_i^{0}, i=1,\ldots,2K$. Select an accuracy level $\epsilon >0$. Let $n=0$.

{\bf Repeat}:\\
    1). Given $\boldsymbol{u}_i^{n}, \boldsymbol{A}^{n}$, update the receive combining vectors $\boldsymbol{v}_i^{n+1}, i=1,\ldots,2K$, via (\ref{eq.optvi}). \\
    2). With $\boldsymbol{u}_i^{n} $ and $\boldsymbol{v}_i^{n+1}$ fixed, use Algorithm 1 to solve the max-min SINR problem (\ref{eq.maxminRelaymimo}) to obtain the relay beamforming matrix $\boldsymbol{A}^{n+1}$.\\
    3).  With $\boldsymbol{A}^{n+1}$ and $\boldsymbol{v}_i^{n+1}$ fixed, solve the max-min SINR problem (\ref{eq.relaxui}) to compute its optimal value $\lambda_{\boldsymbol{u}}^{n}$ and the corresponding users' transmit precoding vectors $\boldsymbol{u}_i^{n+1}, i=1,\ldots,2K$, via Algorithm 1 (with minor modification).\\
    4). Set $n=n+1$.\\
{\bf until} $ |\lambda_{\boldsymbol{u}}^{n} - \lambda_{\boldsymbol{u}}^{n-1}| < \epsilon$.

\end{algorithm}
\vspace{0.05 in} \hrule height0.1pt depth0.3pt \vspace{0.1 in}
\end{quote}

Since the objective of the intended problem (\ref{eq.maxmin2kmimo}) is clearly upper-bounded and it is increased in each iteration of Algorithm 3, the convergence of the proposed alternative optimization approach readily follows. Note that Algorithm 3 in general converges to a local optimum point. Nevertheless, as will be shown in the next section, the beamforming design with the proposed iterative algorithm can significantly outperform the
existing methods.

For weighted sum-rate maximization and other criteria, a similar iterative optimization algorithm can be developed to find the users' transmit precoding vectors, the relay's beamforming matrix, and the users' receive combining vectors. Consider the beamforming designs for weighted sum-rate maximization. The joint design problem can be again decoupled into three sub-problems
and an iterative method can be used to alternatively solve
the three sub-problems. Specifically, during the $n$-th iteration, we first update the users' receive combining vectors $\boldsymbol{v}_i^{n+1}, i=1,\ldots,2K$, via (\ref{eq.optvi}) with fixed  $\boldsymbol{u}_i^{n}, i=1,\ldots,2K$, and $\boldsymbol{A}^{n}$. Given $\boldsymbol{u}_i^{n} $ and $\boldsymbol{v}_i^{n+1}$, we next find the optimal relay beamforming matrix $\boldsymbol{A}^{n+1}$. This sub-optimization problem is an MP. Building on the max-min SINR solution to (\ref{eq.maxminRelaymimo}),  Algorithm 2 can be used to obtain $\boldsymbol{A}^{n+1}$. With $\boldsymbol{A}^{n+1}$ and $\boldsymbol{v}_i^{n+1}$ fixed, the optimal precoding vectors $\boldsymbol{u}_i^{n+1}, i=1,\ldots,2K$, for weighted sum-rate maximization can also be found by the polyblock outer approximation method in Algorithm 2 building on the max-min SINR solution to (\ref{eq.relaxui}). It is guaranteed that the proposed MP based alternative optimization approach converges to, at least, a local optimum.

\section{Numerical Results}
In this section, numerical results are presented to test the
proposed beamforming designs. The simulation settings are as follows. We consider uncorrelated Rayleigh flat
fading channels, i.e., each element in $\boldsymbol{h}_i$  or
$\boldsymbol{H}_i$ is independent complex Gaussian distributed with
zero mean and unit variance. Unless otherwise specified, each user
is equipped with a single antenna; the noise components are complex
white Gaussian with $ \boldsymbol{n}_R(t) \thicksim {\cal
CN}(0,N_0\boldsymbol{I}_M)$, and $n_i(t) \thicksim
{\cal CN}(0,N_0)$; assume $p_i=p, \forall i$, and define
$SNR=p/N_0$.
%

\subsection{One-pair TWR}

\begin{figure}[t]
\vspace{-0.2cm} \centering \epsfig{file=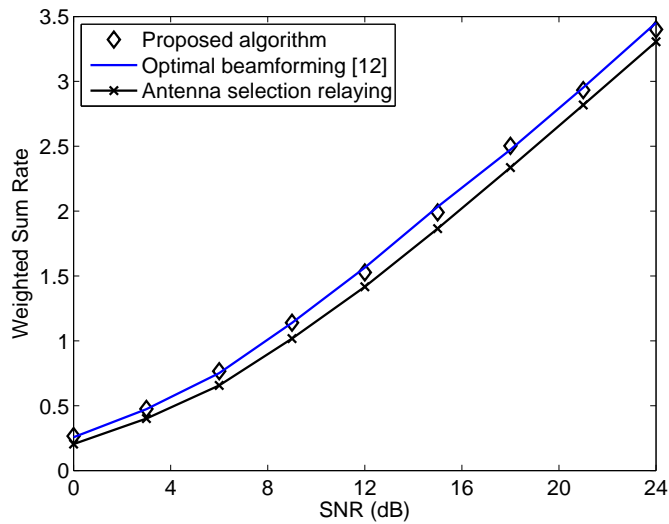,
width=0.55\textwidth} \caption{ Weighted sum-rate of two-user
two-way relaying with various schemes, $p_1=p_2=\check{P}_R,
w_1=0.2,w_2=0.8$, and $M=2$.} \label{F:fig3} \vspace{-0.3cm}
\end{figure}


\begin{figure}[t]
\vspace{-0.2cm} \centering \epsfig{file=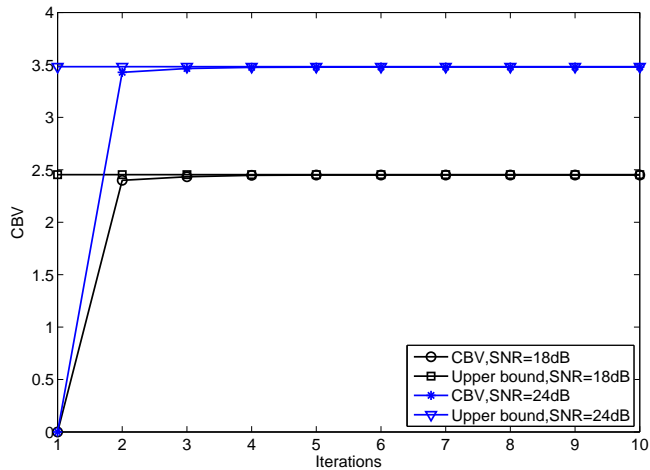,
width=0.55\textwidth} \caption{Evolution of CBV in Algorithm 2.}
\label{F:fig4} \vspace{-0.3cm}
\end{figure}

In Fig. \ref{F:fig3}, we check the optimality of the proposed
monotonic program based weighted sum-rate maximization beamforming
design method for $K=1$ user pair, by comparing with the optimal
beamforming scheme in \cite{Zha09}, and the antenna selection
relaying scheme, where the best antenna is selected for signal
relaying. There are $M=2$ antennas at the relay, and the transmit
power of the relay and the two users are the same :
$p_1=p_2=\check{P}_R$. The weights are chosen as $w_1=0.2$ and
$w_2=0.8$, and $\epsilon =0.01$ for Algorithm 2. It is
seen that the proposed monotonic program based design method
achieves the same performance as the scheme in \cite{Zha09}, which
confirms that the beamforming matrix obtained by Algorithm 2 is
optimal. (The slight differences between the two are due to numerical errors.) To illustrate the convergence behavior of the proposed
method, the CBV in Algorithm 2 is shown in Fig. \ref{F:fig4}. The
weighted sum-rate upper bound is obtained as follows: we ignore the
rank-one constrain when solving the problem (\ref{eq.pt}), and
find the minimal of $\Phi(\boldsymbol{z}^n)$ in Algorithm 2 as
the upper bound. We see that Algorithm 2
converges fast. In this particular example, three iterations is sufficient to determine the
optimal beamforming matrix.

\subsection{Multi-pair TWR}
\begin{figure}
\centering \epsfig{file=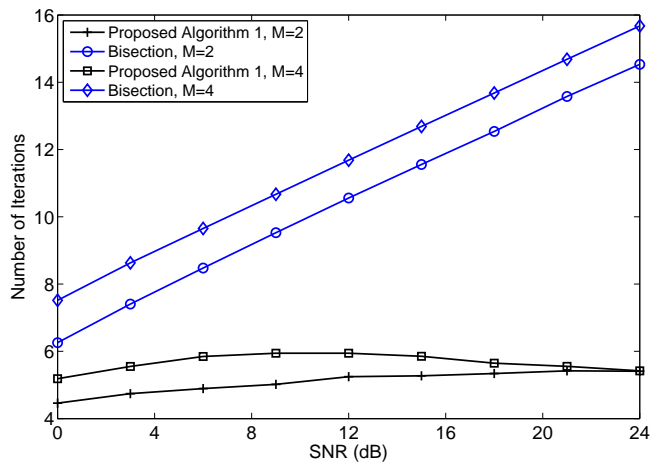, width=0.55\textwidth} \caption{
Comparison between the proposed Algorithm 1 and the bisection search
method in \cite{Tao12} for $\epsilon =0.01$ and $ K=2$. }
\label{F:fig5} \vspace{-0.3cm}
\end{figure}

Now consider a two-pair TWR with a four-antenna relay, i.e., $K = 2$ and $M = 4$. We assume
equal power allocation among the four users and the relay. Fig.
\ref{F:fig5} compares the number of iterations of the proposed
Dinkelbach-type Algorithm 1 with the bisection search method in
\cite{Tao12} for a given solution accuracy $\epsilon=0.01$. For the
bisection method in \cite{Tao12} , the number of iterations is $
\left \lceil \log_2(t/\epsilon) \right \rceil$, where $t$ and
$\epsilon$ are the search bound and error precision, respectively.
The search bound $t$ depends on the SNR and the channel coefficients
\cite{Tao12}. Hence, the number of iterations of the bisection
method increases as the SNR increases or the number of antennas
increases as shown in the figure. On the other hand, the number of
iterations for the proposed Dinkelbach-type Algorithm 1 remains almost unchanged. Using the
zero-forcing beamforming matrix in \cite{Jou10} as the initial
$\boldsymbol{A}^0$, it can be seen that the
proposed Algorithm 1 converges much faster than the
bisection method. About 5 or 6 iterations are sufficient for the convergence of
Algorithm 1 in the whole SNR region.

\begin{figure}
\centering \epsfig{file=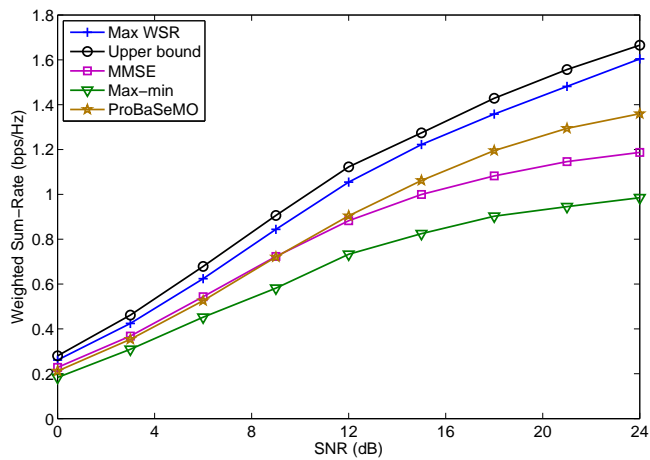, width=0.55\textwidth} \caption{
Weighted sum-rate of four-user TWR with different beamforming
schemes, $w_1=0.2$, $w_2=0.8$, $w_3=0.5$, $w_4=0.5$, and $M=2$.}
\label{F:fig6} \vspace{-0.3cm}
\end{figure}

\begin{figure}
\centering \epsfig{file=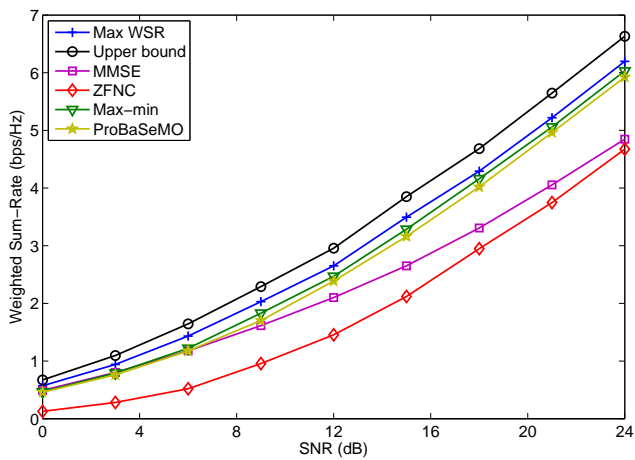, width=0.55\textwidth} \caption{
Weighted sum-rate of four-user TWR with different beamforming
schemes, $w_1=0.2$, $w_2=0.8$, $w_3=0.5$ ,$w_4=0.5$, and $M=4$.}
\label{F:fig7} \vspace{-0.3cm}
\end{figure}

Fig. \ref{F:fig6} and Fig. \ref{F:fig7} show the achievable weighted
sum-rate of various beamforming schemes with $M=2$ and 4 antennas at
the relay, respectively. The weights are chosen as $w_1=0.2$,
$w_2=0.8$, and $w_3=w_4=0.5$. For the proposed weighted sum-rate
maximization (Max WSR) beamforming, the optimal beamforming matrix
$\boldsymbol{A}^{\text{opt}}$ is obtained by the monotonic program
method in Algorithm 2 with $\epsilon = 0.01$. The  weighted sum-rate performance upper bound is obtained as in Fig. \ref{F:fig4}. We compare the
proposed design with the following methods: 1) max-min beamforming
in \cite{Tao12}, 2) minimum mean-square-error (MMSE) beamforming in
\cite{Jou10}, 3) zero-forcing based network coding (ZFNC) in
\cite{WangF12}, and 4) ProBaSeMO scheme in \cite{ZRH12}. Note that
for the ZFNC scheme, the number of antennas at the relay should be
no less than the number of users, hence it is only applicable when
$M=4$. From both figures, it is shown that the performance of the
proposed beamforming design is close to the performance upper bound, and it outperforms all other alternatives for all SNR values.
In particular, the MP approach building on the max-min SINR solution can
significantly improve the sum-rate performance, when there
is only two antennas at the relay.

\begin{figure}
\centering \epsfig{file=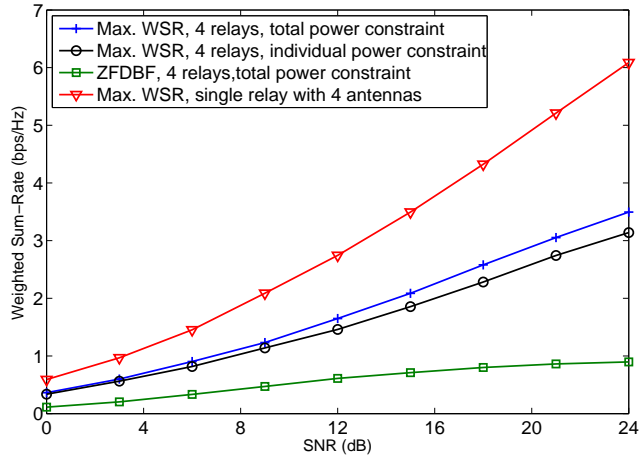, width=0.55\textwidth} \caption{
Weighted sum-rate of multi-pair collaborative TWR with $p_i =
\check{P}_R, \forall i$.} \label{F:fig10} \vspace{-0.3cm}
\end{figure}

\subsection{Collaborative Multi-pair TWR}

Now consider a collaborative four-user TWR with four single-antenna
relays. Fig. \ref{F:fig10} shows the performance of the proposed
collaborative beamforming design and the zero-forcing distributed
beamforming (ZFDBF) scheme in \cite{Wang11}. The simulation
parameters are the same as in Fig. \ref{F:fig6}. We consider two
transmit power constraints: 1) the relays have a total transmit
power constraint that $\sum_{m=1}^{M} \check{P}_{R_m} = p$,
and 2) each relay has individual transmit power constraint
that $\check{P}_{R_m} = p/M, \forall m$. For the considered two
transmit power constraints, it is shown that the collaborative TWR with
total transmit power constraint slightly outperforms that with individual transmit power constraint in the high SNR region.
Compared with the ZFDBF scheme, significant performance gains can be
achieved with the proposed beamforming designs. It can be also seen
that the achievable weighted sum-rate of collaborative TWR with four
single-antenna relays is much lower than that of TWR with a single
four-antenna relay. This is due to the fact that the beamforming
matrix $\boldsymbol{\tilde{A}}$ for collaborative TWR is restricted
to be diagonal. Hence certain multiplexing gain is lost as compared
with the single multi-antenna relay case.

\subsection{MIMO Multi-pair TWR}

\begin{figure}
\centering \epsfig{file=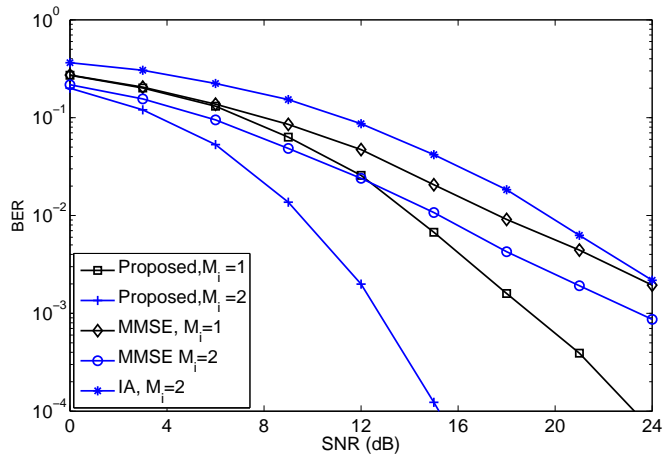, width=0.55\textwidth} \caption{
Average BER of multiuser MIMO TWR with the proposed beamforming
design. $p_i = \check{P}_R , \forall i$ and $M_1 = M_2 = ... =
M_{2K}$.} \label{F:fig11} \vspace{-0.3cm}
\end{figure}

Finally, Fig. \ref{F:fig11} presents the BER performance of a
four-user MIMO TWR system with QPSK modulation, where both the users
and the relay are equipped with multiple antennas. The number of
antennas for one user varies from 1 to 2, and there are 4 antennas
at the relay. It is shown that the BER performance
improves as the number of antennas at each user increases. Also,
significant performance improvement is observed for the proposed
optimal beamforming as compared with the MMSE beamforming scheme in
\cite{Jou10} and the interference alignment (IA) scheme in
\cite{Ganesan11}. For instance, there is more than 10dB gain at a
BER of $10^{-3}$ for the proposed design when there are two antennas
at each user.

\section{Conclusion}
We developed a unified framework of beamforming designs for non-regenerative two-way relaying. Using the max-min SINR
solution as a corner stone, we proposed efficient algorithms to find
the near-optimal beamforming designs under various important criteria
such as power minimization, rate maximization, MSE minimization, and
BER minimization. We further extended the proposed framework to
distributed beamforming for TWR, as well as to MIMO TWR. The
proposed unified approach can provide important insights for
tackling the optimal beamforming designs in other emerging network
models and settings.

\section*{Appendix}

We first show that
\begin{lemma}
$\tilde{\lambda}^{\text{opt}}(\check{P}_R)$ is a strictly increasing function of $\check{P}_R$.
\end{lemma}

{\em Proof}:
Let $\boldsymbol{X}^{\text{opt}}$ denote the optimal solution for (\ref{eq.relax}) with power budget $\check{P}_R>0$. For a $\check{P}_R' > \check{P}_R$, let $\alpha = \check{P}_R' / \check{P}_R >1$, and $\boldsymbol{X}' = \alpha \boldsymbol{X}^{\text{opt}}$. Then $\boldsymbol{X}'$ is feasible for (\ref{eq.relax}) with power budget $\check{P}_R'$, since $\text{tr}(\boldsymbol{E}_0 \boldsymbol{X}') = \alpha \text{tr}(\boldsymbol{E}_0 \boldsymbol{X}^{\text{opt}}) \leq \alpha \check{P}_R = \check{P}_R'$.

On the other hand,
\begin{align}
    \nonumber & {\text{SINR}_i(\boldsymbol{X}') = \frac{f_i(\boldsymbol{X}')}{g_i(\boldsymbol{X}')} = \frac{\text{tr}(\boldsymbol{E}_i^{(1)} \boldsymbol{X}')}{\text{tr}(\boldsymbol{E}_i^{(2)}\boldsymbol{X}') + \sigma_i^2} } = \frac{\alpha \text{tr}(\boldsymbol{E}_i^{(1)} \boldsymbol{X}^{\text{opt}})}{\alpha \text{tr}(\boldsymbol{E}_i^{(2)}\boldsymbol{X}^{\text{opt}}) + \sigma_i^2}\\
    \nonumber & ~~~~~~~  > \frac{ \text{tr}(\boldsymbol{E}_i^{(1)} \boldsymbol{X}^{\text{opt}})}{\text{tr}(\boldsymbol{E}_i^{(2)}\boldsymbol{X}^{\text{opt}}) +
    \sigma_i^2} = \text{SINR}_i(\boldsymbol{X}^{\text{opt}}).
\end{align}
Therefore, $\tilde{\lambda}^{\text{opt}}(\check{P}_R') \geq \min_{i=1,\ldots, 2K} \frac{\text{SINR}_i(\boldsymbol{X}')}{\gamma_i} > \min_{i=1,\ldots, 2K} \frac{\text{SINR}_i(\boldsymbol{X}^{\text{opt}})}{\gamma_i} = \tilde{\lambda}^{\text{opt}}(\check{P}_R)$. \hfill $\square$

Relying on the monotonicity of $\tilde{\lambda}^{\text{opt}}(\check{P}_R)$ stated in Lemma 3, we can further show that:

\begin{lemma}
The optimal solution for (\ref{eq.power33}) is the same as the matrix $\boldsymbol{X}^{\text{opt}}$ for (\ref{eq.relax}) with the power budget $P_R$ that satisfies $\tilde{\lambda}^{\text{opt}}(P_R) = 1$.
\end{lemma}

{\em Proof }:
Let $\boldsymbol{X}^{\text{opt}}$ denote the optimal solution for (\ref{eq.relax}) with the power budget $P_R$  that satisfies $\tilde{\lambda}^{\text{opt}}(P_R) = 1$. 
Since $\tilde{\lambda}^{\text{opt}}(P_R) = 1$ implies $\text{SINR}_i(\boldsymbol{X}^{\text{opt}}) \geq \gamma_i$, $i=1,\ldots, 2K$, $\boldsymbol{X}^{\text{opt}}$ is in the feasible set of (\ref{eq.power33}). Upon denoting $P_R^{\text{opt}}$ as the optimal value for (\ref{eq.power33}), this in turn implies that $P_R^{\text{opt}} \leq \text{tr}(\boldsymbol{E}_0 \boldsymbol{X}^{\text{opt}}) \leq P_R$. Consider (\ref{eq.relax}) with the power budget $P_R^{\text{opt}}$. By Lemma 3, we must have
\begin{equation}\label{eq.lemma31}
    \tilde{\lambda}^{\text{opt}}(P_R^{\text{opt}}) \leq \tilde{\lambda}^{\text{opt}}(P_R) = 1
\end{equation}
due to $P_R^{\text{opt}} \leq P_R$.

On the other hand, let $\boldsymbol{\tilde{X}}^{\text{opt}}$ denote the optimal solution for (\ref{eq.power33}), which is the feasible set of (\ref{eq.relax}) with the power budget $P_R^{\text{opt}}$ since $\text{tr}(\boldsymbol{E}_0 \boldsymbol{\tilde{X}}^{\text{opt}}) =P_R^{\text{opt}}$. For this $\boldsymbol{\tilde{X}}^{\text{opt}}$, we have $\min_{i=1, \ldots,2K} \frac{\text{SINR}_i(\boldsymbol{\tilde{X}}^{\text{opt}})}{\gamma_i} \geq 1$ since $\text{SINR}_i(\boldsymbol{\tilde{X}}^{\text{opt}}) \geq \gamma_i$, $i=1,\ldots, 2K$. This together with the feasibility of $\boldsymbol{\tilde{X}}^{\text{opt}}$ implies that $\tilde{\lambda}^{\text{opt}}(P_R^{\text{opt}}) \geq 1$. Clearly, we have both the latter and (\ref{eq.lemma31}) satisfied, only when all the inequalities are satisfied with equalities; i.e., $P_R^{\text{opt}}= P_R$, and it is achieved by the beamforming matrix $\boldsymbol{X}^{\text{opt}}$. \hfill $\square$

\end{document}